\newcommand{\Ntotal}  {877~}
\newcommand{\intLdt}  {57~}
\newcommand{\intLdtfull}  {57.21}
\newcommand{\dLtot}   {0.25}
\newcommand{\rroots}  {182.68}  
\newcommand{\droots}  {0.05} 

\newcommand{\GENTxs}   {15.72}
\newcommand{\measxs}  {15.43}
\newcommand{\sigst}  {0.61}
\newcommand{\sigsys}  {0.26}
\newcommand{\measbr}  {67.9}
\newcommand{\brst}  {1.2}
\newcommand{\brsys}  {0.5}

\newcommand{\resdkg} {\mbox{$0.11^{+0.52}_{-0.37}$}} 
\newcommand{\reslam} {\mbox{$-0.10^{+0.13}_{-0.12}$}}
\newcommand{\resdgz} {\mbox{$0.01^{+0.13}_{-0.12}$}}
\newcommand{\resawf} {\mbox{$-0.04 \pm 0.09$}}

\newcommand{\abftyr} {$ -0.08^{+0.71}_{-0.54}$}
\newcommand{\awtyr}  {$ -0.16 \pm 0.16$}
\newcommand{\dgztyr} {$ -0.01 \pm 0.15$}

\newcommand{\abfbot} {$  0.04^{+0.67}_{-0.44}$} 
\newcommand{\awbot}  {$ -0.14^{+0.15}_{-0.14}$}
\newcommand{\dgzbot} {$ -0.01 \pm 0.14$}

\newcommand{\dabfhe}    {\mbox{$  .088$}}
\newcommand{\dabfmc}    {\mbox{$  .049$}}

\newcommand{\dabfsu}    {\mbox{$  .178$}}
\newcommand{\abfsta}    {\mbox{$ -0.21^{+0.61}_{-0.45}$}}
\newcommand{\abfold}    {\mbox{$ -0.18^{+0.77}_{-0.56}$}}

\newcommand{\abfexp}    {\mbox{$\pm 0.46$}}

\newcommand{ \dawhe}    {\mbox{$  .012$}}
\newcommand{ \dawmc}    {\mbox{$  .011$}}

\newcommand{ \dawsu}    {\mbox{$  .038$}}
\newcommand{ \awsta}    {\mbox{$ -0.19 \pm 0.15$}}
\newcommand{ \awold}    {\mbox{$ -0.19 \pm 0.16$}}

\newcommand{ \awexp}    {\mbox{$\pm 0.14$}}

\newcommand{\ddgzhe}    {\mbox{$  .012$}}
\newcommand{\ddgzmc}    {\mbox{$  .003$}}

\newcommand{\ddgzsu}    {\mbox{$  .043$}}
\newcommand{\dgzsta}    {\mbox{$  0.00 \pm 0.14$}}
\newcommand{\dgzold}    {\mbox{$  0.00 \pm 0.16$}}

\newcommand{\dgzexp}    {\mbox{$\pm 0.13$}}
\newcommand{\Opal}{\mbox{OPAL}}
\newcommand{\LepII}{\mbox{LEP}}
\newcommand{\LepI}{\mbox{LEP}}
\newcommand{\Jetset}{\mbox{J{\sc etset}}}
\newcommand{\Pythia}{\mbox{P{\sc ythia}}}
\newcommand{\Herwig}{\mbox{H{\sc erwig}}}
\newcommand{\Ariadne}{\mbox{A{\sc riadne}}}
\newcommand{\Excalibur}{\mbox{E{\sc xcalibur}}}
\newcommand{\Gentle}{\mbox{G{\sc entle}}}
\newcommand{\Koralw}{\mbox{K{\sc oralw}}}
\newcommand{\Koralz}{\mbox{K{\sc oralz}}}
\newcommand{\Bhwide}{\mbox{B{\sc hwide}}}
\newcommand{\Phojet}{\mbox{P{\sc hojet}}}

\newcommand{\Com}{centre-of-mass}
\newcommand{\Grace}{\mbox{\tt grc4f}}
\newcommand{\SM}{Standard Model}
\newcommand{\MC}{Monte Carlo}
\newcommand{\tgc}{{\small TGC}}

\newcommand{\GeV}{\mbox{$\mathrm{GeV}$}}

\newcommand{\GeVcc}{\mbox{$\mathrm{GeV}\!/\!{\it c}^2$}}
\newcommand{\Ipb}{\mbox{pb$^{-1}$}}
\newcommand{\beq}{\begin{equation}}
\newcommand{\eeq}{\end{equation}}
\newcommand{\bea}{\begin{eqnarray}}
\newcommand{\eea}{\end{eqnarray}}
\newcommand{\ra}{\mbox{$\rightarrow$}}

\def\gappeq{\mathrel{ \rlap{\raise.5ex\hbox{$>$}}
                      {\lower.5ex\hbox{$\sim$}}  } }
\def\lappeq{\mathrel{ \rlap{\raise.5ex\hbox{$<$}}
                      {\lower.5ex\hbox{$\sim$}}  } }
\newcommand{\etal}{{\it et al.}}

\newcommand{\Mw}{\mbox{$M_{\mathrm{W}}$}}

\newcommand{\cwsq}{\mbox{$\cos^2\theta_w$}}
\newcommand{\twsq}{\mbox{$\tan^2\theta_w$}}
\newcommand{\Vij} {\mbox{$|\mathrm{V}_{ij}|$}}
\newcommand{\Vud} {\mbox{$|\mathrm{V}_{\mathrm{ud}}|$}}
\newcommand{\Vus} {\mbox{$|\mathrm{V}_{\mathrm{us}}|$}}
\newcommand{\Vcd} {\mbox{$|\mathrm{V}_{\mathrm{cd}}|$}}
\newcommand{\Vcb} {\mbox{$|\mathrm{V}_{\mathrm{cb}}|$}}
\newcommand{\Vub} {\mbox{$|\mathrm{V}_{\mathrm{ub}}|$}}
\newcommand{\Vcs} {\mbox{$|\mathrm{V}_{\mathrm{cs}}|$}}
\newcommand{\Br}{\mbox{$\mathrm{Br}$}}
\newcommand{\epem}{\mbox{$\mathrm{e^+e^-}$}}

\newcommand{\lplm}{\mbox{$\ell\overline{\ell}$}}
\newcommand{\Zz}{\mbox{${\mathrm{Z}^0}$}}
\newcommand{\WW}{\mbox{$\mathrm{W^+W^-}$}}
\newcommand{\Wm}{\mbox{$\mathrm{W^-}$}}
\newcommand{\Wp}{\mbox{$\mathrm{W^+}$}}
\newcommand{\WT}{\mbox{$\mathrm{W_T}$}}
\newcommand{\WL}{\mbox{$\mathrm{W_L}$}}

\newcommand{\eeWW}{\mbox{\epem$\rightarrow$\WW}}

\newcommand{\qq}{\mbox{$\mathrm{q\overline{q}}$}}
\newcommand{\Qqll}{\qq\lplm}
\newcommand{\Qqnn}{\qq\nn}
\newcommand{\Llll}{\lplm\lplm}
\newcommand{\Llnn}{\lplm\nn}
\newcommand{\qqgg}{\mbox{$\qq{\mathrm{gg}}$}}
\newcommand{\qqqq}{\mbox{$\qq\qq$}}
\newcommand{\Wqq}{\mbox{$\mathrm{q\overline{q} }$}}
\newcommand{\pbp}{\mbox{$\mathrm{\overline{p}p }$}}

\newcommand{\lnu}{\mbox{$\overline{\ell}\nu_{\ell}$}}
\newcommand{\lnubar}{\mbox{$\ell^\prime\overline{\nu}_{\ell^\prime}$}}

\newcommand{\enu}{\mbox{$\mathrm{e\overline{\nu}_{e}}$}}
\newcommand{\mnu}{\mbox{$\mu\overline{\nu}_{\mu}$}}
\newcommand{\tnu}{\mbox{$\tau\overline{\nu}_{\tau}$}}
\newcommand{\emnu}{\mbox{$\mathrm{e^-\overline{\nu}_{e}}$}}
\newcommand{\mmnu}{\mbox{$\mu^-\overline{\nu}_{\mu}$}}
\newcommand{\tmnu}{\mbox{$\tau^-\overline{\nu}_{\tau}$}}
\newcommand{\epnu}{\mbox{$\mathrm{e^+{\nu_{e}}}$}}
\newcommand{\mpnu}{\mbox{$\mu^+{\nu}_{\mu}$}}
\newcommand{\tpnu}{\mbox{$\tau^+{\nu}_{\tau}$}}

\newcommand{\Wtnu}{\mbox{$\Wm\ra\tau\overline{\nu}_{\tau}$}}
\newcommand{\WWqqqq}{\mbox{\WW$\rightarrow$ \Wqq\Wqq}}
\newcommand{\WWqqln}{\mbox{\WW$\rightarrow$ \Wqq\lnu}}
\newcommand{\WWqqen}{\mbox{\WW$\rightarrow$ \Wqq\enu}}
\newcommand{\WWqqmn}{\mbox{\WW$\rightarrow$ \Wqq\mnu}}
\newcommand{\WWqqtn}{\mbox{\WW$\rightarrow$ \Wqq\tnu}}
\newcommand{\WWlnln}{\mbox{\WW$\rightarrow$ \lnu\lnubar}}
\newcommand{\WWenen}{\mbox{\WW$\rightarrow$ \epnu\emnu}}
\newcommand{\WWenmn}{\mbox{\WW$\rightarrow$ \epnu\mmnu}}
\newcommand{\WWentn}{\mbox{\WW$\rightarrow$ \epnu\tmnu}}
\newcommand{\WWmnmn}{\mbox{\WW$\rightarrow$ \mpnu\mmnu}}
\newcommand{\WWmntn}{\mbox{\WW$\rightarrow$ \mpnu\tmnu}}
\newcommand{\WWtntn}{\mbox{\WW$\rightarrow$ \tpnu\tmnu}}
\newcommand{\Wenu}{\mbox{$\epem \rightarrow \mathrm{W}\enu$}}
\newcommand{\Wqqp}{\mbox{$\mathrm{W}\rightarrow\qq$}}

\newcommand{\Zee}{\mbox{$\epem\rightarrow\Zz\epem$}}
\newcommand{\ZZ}{\mbox{$\epem\rightarrow\Zz\Zz$}}
\newcommand{\Zqq}{\mbox{$\Zz\rightarrow\qq$}}

\newcommand{\ZGqq}{\mbox{$\Zz/\gamma\rightarrow\qq$}}
\newcommand{\ZGll}{\mbox{$\Zz/\gamma\rightarrow\lplm$}}

\newcommand{\roots}{\mbox{$\sqrt{s}$}}

\newcommand{\Zgamma}{\mbox{$\Zz/\gamma$}}

\newcommand{\sigccthree}{\mbox{$\sigma_{\mathrm{WW}}$}}

\newcommand{\al}{\mbox{$\alpha$}}
\newcommand {\ee} {\mbox{$\mathrm{e}^+ \mathrm{e}^-$}}

\newcommand {\mm} {\mbox{$\mu^+ \mu^-$}}
\newcommand {\nn} {\mbox{$\nu\overline{\nu}$}}
\newcommand {\tautau} {\mbox{$\tau^+ \tau^-$}}

\newcommand {\eeee}   {\ee\ra\ee}
\newcommand {\eemumu} {\ee\ra\mm}
\newcommand {\eetautau} {\ee\ra\tautau}
\newcommand {\eenunu} {\ee\ra\nn}

\newcommand{\Wmv}{\mbox{$\mathrm{W}\rightarrow\mnu$}}
\newcommand{\Wev}{\mbox{$\mathrm{W}\rightarrow\enu$}}
\newcommand{\Wtv}{\mbox{$\mathrm{W}\rightarrow\tnu$}}
\newcommand{\Wlv}{\mbox{$\mathrm{W}\rightarrow\lnu$}}

\newcommand{\Cthw}{\mbox{$\cos\theta_{\mathrm{W}} $}}

\newcommand{\Cthstl}{\mbox{$\cos \theta_\ell^{*}$}}
\newcommand{\Phistl}{\mbox{$\phi_\ell^{*}$}}
\newcommand{\Cthstj}
{\mbox{$\cos\theta_{\scriptscriptstyle \mathrm{jet}}^{*}$}}
\newcommand{\Phistj}{\mbox{$\phi_{\scriptscriptstyle \mathrm{jet}}^{*}$}}

\newcommand{\OO}{\mbox{${\cal O}$}}

\newcommand{\Lnln}{\lnu\lnubar}
\newcommand{\Qqln}{\Wqq\lnu}
\newcommand{\Qqen}{\Wqq\enu}
\newcommand{\Qqmn}{\Wqq\mnu}
\newcommand{\Qqtn}{\Wqq\tnu}
\newcommand{\Qqqq}{\Wqq\Wqq}
\newcommand{\Qqee}{\Wqq\ee}
\newcommand{\semnu}{\mbox{$\mathrm{e^{^-}\!\!\overline{\nu}_{e}}$}}
\newcommand{\smmnu}{\mbox{$\mu^{^-}\!\!\overline{\nu}_{\mu}$}}
\newcommand{\stmnu}{\mbox{$\tau^{^-}\!\!\overline{\nu}_{\tau}$}}
\newcommand{\sepnu}{\mbox{$\mathrm{e^{^+}\!\!{\nu_{e}}}$}}
\newcommand{\smpnu}{\mbox{$\mu^{^+}\!\!{\nu}_{\mu}$}}
\newcommand{\stpnu}{\mbox{$\tau^{^+}\!\!{\nu}_{\tau}$}}
\newcommand{\senen}{\sepnu\semnu}
\newcommand{\senmn}{\sepnu\smmnu}
\newcommand{\sentn}{\sepnu\stmnu}
\newcommand{\smnmn}{\smpnu\smmnu}
\newcommand{\smntn}{\smpnu\stmnu}
\newcommand{\stntn}{\stpnu\stmnu}

\newcommand{\gz}{\mbox{$g_1^{\mathrm{z}}$}}
\newcommand{\kg}{\mbox{$\kappa_\gamma$}}
\newcommand{\kz}{\mbox{$\kappa_{\mathrm{z}}$}}
\newcommand{\lgg}{\mbox{$\lambda_\gamma$}}
\renewcommand{\lg}{\mbox{$\lambda$}}
\newcommand{\lz}{\mbox{$\lambda_{\mathrm{z}}$}}
\newcommand{\dgz}{\mbox{$\Delta g_1^{\mathrm{z}}$}}

\newcommand{\dkg}{\mbox{$\Delta \kappa_\gamma$}}

\newcommand{\dkz}{\mbox{$\Delta \kappa_{\mathrm{z}}$}}

\newcommand{\abf}{\mbox{$\alpha_{B\phi}$}}
\newcommand{\awf}{\mbox{$\alpha_{W\phi}$}}
\newcommand{\aw}{\mbox{$\alpha_W$}}

\newcommand{\LL}{\mbox{$\log L$}}

\newcommand {\dmes}{ density matrix elements}
\newcommand {\sdme}{spin density matrix element}
\newcommand {\sdmes}{spin density matrix elements}
\newcommand {\xse}{cross-section}
\newcommand {\xses}{cross-sections}

\newcommand{\x}{\mbox{$\Cthw$}}
\newcommand{\y}{\mbox{$\cos\theta^*$}}
\newcommand{\z}{\mbox{$\phi^*$}}
\newcommand{\dreidiff}{\mbox{$d\x d\y d\z$}}

\documentstyle[epsfig,a4p,cite,12pt]{article}

\begin{document}
\bibliographystyle{plain}
\begin{titlepage}
\begin{center}{\large   EUROPEAN LABORATORY FOR PARTICLE PHYSICS
 }\end{center}\bigskip
\begin{flushright}
 CERN-EP/98-167 \\ 
 19th October 1998  \\
\end{flushright}
\bigskip\bigskip\bigskip\bigskip\bigskip
\begin{center}
 {\LARGE\bf \boldmath 
\WW\ production and triple gauge boson couplings at LEP
energies up to 183 GeV}
\end{center}
\bigskip\bigskip
\begin{center}{\Large The OPAL Collaboration}
\end{center}\bigskip
\bigskip
\bigskip\begin{center}{\large  Abstract}\end{center}
A study of W-pair production in \epem\ annihilations at \LepII\ is 
presented, based on \Ntotal \WW\ candidates corresponding
to an integrated luminosity of \intLdt\Ipb\ at \roots=183~\GeV.
Assuming that the angular distributions of the W-pair
production and decay, as well as their branching fractions, are 
described by the Standard Model, the W-pair production \xse\
is measured to be \measxs$\pm$\sigst (stat.)$\pm$\sigsys (syst.) pb.  
Assuming lepton universality and combining with our results from lower
centre-of-mass energies, the W branching fraction to hadrons 
is determined to be \measbr$\pm$\brst (stat.)$\pm$\brsys (syst.)\%.
The number of W-pair candidates and the angular
distributions for each final state (\Qqln, \Qqqq, \Lnln)
are used to determine the triple gauge boson couplings.
After combining these values with our results from
lower centre-of-mass energies 
we obtain \dkg=\resdkg, \dgz=\resdgz\ and \lg=\reslam, where the errors
include both statistical and systematic uncertainties and each coupling
is determined setting the other two couplings to the
\SM\ value. The fraction
of W bosons produced with a longitudinal polarisation is measured to be
0.242$\pm$0.091(stat.)$\pm$0.023(syst.). All these 
measurements are consistent with the \SM\ expectations. 
\bigskip\bigskip\bigskip\bigskip
\begin{center}
{\large Submitted to the European Physical Journal C}
\end{center}

\end{titlepage}


\begin{center}{\Large        The OPAL Collaboration
}\end{center}\bigskip
\begin{center}{
G.\thinspace Abbiendi$^{  2}$,
K.\thinspace Ackerstaff$^{  8}$,
G.\thinspace Alexander$^{ 23}$,
J.\thinspace Allison$^{ 16}$,
N.\thinspace Altekamp$^{  5}$,
K.J.\thinspace Anderson$^{  9}$,
S.\thinspace Anderson$^{ 12}$,
S.\thinspace Arcelli$^{ 17}$,
S.\thinspace Asai$^{ 24}$,
S.F.\thinspace Ashby$^{  1}$
% ...
}\end{center}\bigskip
\bigskip
$^{  1}$School of Physics and Astronomy, University of Birmingham,
Birmingham B15 2TT, UK
\newline
$^{  2}$Dipartimento di Fisica dell' Universit\`a di Bologna and INFN,
I-40126 Bologna, Italy
\newline
$^{  3}$Physikalisches Institut, Universit\"at Bonn,
D-53115 Bonn, Germany
\newline
$^{  4}$Department of Physics, University of California,
Riverside CA 92521, USA
\newline
$^{  5}$Cavendish Laboratory, Cambridge CB3 0HE, UK
\newline
$^{  6}$Ottawa-Carleton Institute for Physics,
Department of Physics, Carleton University,
Ottawa, Ontario K1S 5B6, Canada
\newline
$^{  7}$Centre for Research in Particle Physics,
Carleton University, Ottawa, Ontario K1S 5B6, Canada
\newline
$^{  8}$CERN, European Organisation for Particle Physics,
CH-1211 Geneva 23, Switzerland
\newline
$^{  9}$Enrico Fermi Institute and Department of Physics,
University of Chicago, Chicago IL 60637, USA
\newline
$^{ 10}$Fakult\"at f\"ur Physik, Albert Ludwigs Universit\"at,
D-79104 Freiburg, Germany
\newline
$^{ 11}$Physikalisches Institut, Universit\"at
Heidelberg, D-69120 Heidelberg, Germany
\newline
$^{ 12}$Indiana University, Department of Physics,
Swain Hall West 117, Bloomington IN 47405, USA
\newline
$^{ 13}$Queen Mary and Westfield College, University of London,
London E1 4NS, UK
\newline
$^{ 14}$Technische Hochschule Aachen, III Physikalisches Institut,
Sommerfeldstrasse 26-28, D-52056 Aachen, Germany
\newline
$^{ 15}$University College London, London WC1E 6BT, UK
\newline
$^{ 16}$Department of Physics, Schuster Laboratory, The University,
Manchester M13 9PL, UK
\newline
$^{ 17}$Department of Physics, University of Maryland,
College Park, MD 20742, USA
\newline
$^{ 18}$Laboratoire de Physique Nucl\'eaire, Universit\'e de Montr\'eal,
Montr\'eal, Quebec H3C 3J7, Canada
\newline
$^{ 19}$University of Oregon, Department of Physics, Eugene
OR 97403, USA
\newline
$^{ 20}$CLRC Rutherford Appleton Laboratory, Chilton,
Didcot, Oxfordshire OX11 0QX, UK
\newline
$^{ 22}$Department of Physics, Technion-Israel Institute of
Technology, Haifa 32000, Israel
\newline
$^{ 23}$Department of Physics and Astronomy, Tel Aviv University,
Tel Aviv 69978, Israel
\newline
$^{ 24}$International Centre for Elementary Particle Physics and
Department of Physics, University of Tokyo, Tokyo 113-0033, and
Kobe University, Kobe 657-8501, Japan
\newline
$^{ 25}$Institute of Physical and Environmental Sciences,
Brunel University, Uxbridge, Middlesex UB8 3PH, UK
\newline
$^{ 26}$Particle Physics Department, Weizmann Institute of Science,
Rehovot 76100, Israel
\newline
$^{ 27}$Universit\"at Hamburg/DESY, II Institut f\"ur Experimental
Physik, Notkestrasse 85, D-22607 Hamburg, Germany
\newline
$^{ 28}$University of Victoria, Department of Physics, P O Box 3055,
Victoria BC V8W 3P6, Canada
\newline
$^{ 29}$University of British Columbia, Department of Physics,
Vancouver BC V6T 1Z1, Canada
\newline
$^{ 30}$University of Alberta,  Department of Physics,
Edmonton AB T6G 2J1, Canada
\newline
$^{ 31}$Duke University, Dept of Physics,
Durham, NC 27708-0305, USA
\newline
$^{ 32}$Research Institute for Particle and Nuclear Physics,
H-1525 Budapest, P O  Box 49, Hungary
\newline
$^{ 33}$Institute of Nuclear Research,
H-4001 Debrecen, P O  Box 51, Hungary
\newline
$^{ 34}$Ludwigs-Maximilians-Universit\"at M\"unchen,
Sektion Physik, Am Coulombwall 1, D-85748 Garching, Germany
\newline
\bigskip\newline
$^{  a}$ and at TRIUMF, Vancouver, Canada V6T 2A3
\newline
$^{  b}$ and Royal Society University Research Fellow
\newline
$^{  c}$ and Institute of Nuclear Research, Debrecen, Hungary
\newline
$^{  d}$ and Department of Experimental Physics, Lajos Kossuth
University, Debrecen, Hungary
\newline
$^{  e}$ on leave of absence from the University of Freiburg
\newline
\newpage


\section{Introduction}
 \label{sec:intro}

One of the main motivations to double the LEP energy was to
study the W boson properties as well as the
characteristics of the W-pair production process, such as the
total \xse, the angular distributions and the 
helicity structure. The total \xse\ for different final states can be 
used to measure the W decay branching fractions in order 
to test lepton universality and to extract information about the 
Cabibbo-Kobayashi-Maskawa quark mixing matrix.

In addition to the $t$-channel $\nu$-exchange,
W-pair production in \epem\ annihilations involves the triple gauge 
boson vertices WW$\gamma$ and WWZ which are present in the \SM\
due to its non-Abelian nature. Therefore, all the W-pair production
properties can be interpreted in terms of the Triple Gauge Couplings (TGCs).
Any deviation from the \SM\ predictions would be evidence for new physics. 

The most general Lorentz invariant 
Lagrangian~\cite{LEP2YR,HAGIWARA,BILENKY,GAEMERS} 
which describes the triple gauge boson interaction involving W bosons
has fourteen independent terms, seven describing  the WW$\gamma$ 
vertex and seven describing the WWZ vertex.  
This parameter space is very large, and it is not currently possible
to measure all fourteen couplings independently. 
Assuming that the Lagrangian satisfies electromagnetic gauge invariance
and charge conjugation as well as parity invariance, 
the number of parameters reduces to five,  which can be taken
as \gz, \kz, \kg, \lz\ and \lgg~\cite{LEP2YR,HAGIWARA}. 
In the \SM\, \gz=\kz=\kg=1 and \lz=\lgg=0.
Considerations related to SU(2)$\times$U(1) gauge invariance, supported
by constraints arising from precise measurements on the \Zz\ resonance
and lower energy data, suggest the following relations between the five
couplings~\cite{LEP2YR,BILENKY},
\begin{eqnarray}
\label{su2u1}
\dkz & = & -\dkg \twsq +  \dgz, \\ \nonumber
\lz  & = &  \lgg, \nonumber
\end{eqnarray} 
where the $\Delta$ indicates a deviation of the respective quantity 
from its \SM\, value and $\theta_w$ is the weak mixing angle.
These constraints leave only three independent couplings\footnote{
In our previous publications we analysed the $\alpha$ parameters 
which are motivated by specific theoretical models and  
are given by, \abf= \dkg -- \dgz \cwsq, \awf= \dgz \cwsq\ and
\aw =\lg. In this paper, however, we prefer to use parameters 
which are directly related to the W electromagnetic and weak 
properties~\cite{HAGIWARA,BILENKY}.},
$\dkg$, $\dgz$ and \lg(=\lgg=\lz) which are not significantly 
restricted~\cite{DERUJULA,HISZ} by existing \LepI\ \Zz\ data.

First TGC studies were made at \pbp\ colliders, mainly at the 
Tevatron by CDF~\cite{CDF} and D0~\cite{D0}, using di-boson production.
Previous results from \LepII\ used the data recorded at
161~\GeV~\cite{tgc161-analysis,OTHERLEP161-tgc} and  
172~\GeV~\cite{tgc172-analysis,OTHERLEP172-tgc} with an integrated 
luminosity of $\approx 10$ \Ipb\ at each centre-of-mass energy.
Here we present an analysis of the OPAL data taken in 1997 at a
centre-of-mass energy of 182.7~\GeV. The TGC results are combined with 
those at lower centre-of-mass energies.

Anomalous \tgc s affect both the total W-pair \xse\ and  
the production angular distribution.
Additionally, the relative contribution of each helicity state  
would be modified, 
which in turn affects the angular distributions of the W decay products.

In the limit of small W width and no initial state radiation (ISR), the
production and decay of W bosons is characterised by five angles.
These are conventionally~\cite{LEP2YR,BILENKY,SEKULIN} taken to be:
the \Wm\ production polar angle\footnote{
The OPAL right-handed coordinate system is defined 
such that the origin is at the geometric centre of the detector,
the $z$-axis is parallel to, and has positive sense, along the e$^-$ beam 
direction, $\theta$ is the polar angle with respect to $z$ and $\phi$ is 
the azimuthal angle around $z$.} $\theta_W$;
the polar and azimuthal angles, $\theta_1^*$ and $\phi_1^*$,
of the decay fermion, \Wm\ra f, in the \Wm\ rest frame\footnote{
The axes of the right-handed coordinate system in the W 
 rest-frame are defined such that $z$ is along
 the parent W flight direction and $y$ is in the direction 
 $\overrightarrow{e^-} \times \overrightarrow{W}$ where 
$\overrightarrow{e^-}$ is the electron beam 
 direction and $\overrightarrow{W}$ is the parent W flight direction.};
and the analogous angles
for the \Wp\ra$\overline{\mathrm{f}}$ decay, $\theta_2^*$ and $\phi_2^*$.
These five angles contain all the information which can be extracted 
about the helicity structure of \WW\ production and decay.
The effects of the finite W width and ISR
are not large at a centre-of-mass energy of around
183~\GeV~\cite{LEP2YR}. However the experimental
accessibility of the different angles in \WW\ events observed in the 
detector, and therefore the sensitivity to
the \tgc s, depends strongly on the final states produced when the W boson
decays.

In this paper we analyse W-pairs in all possible final states, namely,
\WWqqln, where one W boson decays into hadrons and the other decays into
leptons\footnote{
Throughout this paper charge conjugate modes are also included and
\qq\ implicitly means any pair of quark-antiquark of different
(or same) flavour which can be produced in W (\Zz) decay.},
\WWqqqq, where both W bosons decay hadronically, and \WWlnln, where both
W bosons decay into leptons.

Most of the \tgc\ results of this analysis are obtained for each of the 
three couplings separately, setting the other two couplings to zero which 
is the \SM\ value. These results will be presented in tables and 
\LL\ curves\footnote{
Throughout this paper, \LL\ denotes negative log-likelihood.}.
We also perform two-dimensional and three dimensional fits, where
two or all three TGC parameters are allowed to vary in the fits. 
The results of these fits will be presented by contour plots.  

The following section includes a short presentation of the OPAL data and 
Monte Carlo samples, and in section~\ref{sec:evsel} the selection of 
our W-pair sample is described. The analysis of the W-pair \xse\ 
and the W decay branching fractions is presented in section 4, along with 
the interpretation of this analysis in terms of the TGCs. 
The TGC analysis using the angular distributions is done 
separately for each channel and is described in sections \ref{sec:qqln},
\ref{sec:qqqq} and \ref{sec:lnln}. The combined TGC results are presented
in section~\ref{sec:combtgc}. Section~\ref{sec:summary} includes the 
summary and conclusions of this study.


\section{Data and Monte Carlo models}
\label{sec:data}
A detailed description of the \Opal\ detector has been presented 
elsewhere~\cite{OPAL,SW}.
The accepted integrated luminosity in 1997, evaluated 
using small angle Bhabha scattering events observed in the silicon
tungsten forward calorimeter, is 57.21 $\pm$0.15 $\pm$0.20 \Ipb\ where the 
first error is statistical and the second systematic.
The luminosity-weighted mean centre-of-mass energy for the data sample
is \roots=\rroots$\pm$\droots~\GeV. Part of the TGC analysis involves 
the 161 and 172 \GeV\ data samples taken in 1996 at \Com\
energies of 161.33$\pm$0.05 and 172.12$\pm$0.06 \GeV\  with 
corresponding luminosities of 9.89$\pm$0.06 and 10.36$\pm$0.06 \Ipb. 

 The semi-analytic program \Gentle\cite{GENTLE} is used to calculate the
 \WW\ \xse\ for the \SM\ case and also for different values of anomalous
 TGCs. The calculated \SM\ \xse\ is \GENTxs~pb at 
\roots=\rroots~\GeV\ using the W mass of 
\Mw=80.40~\GeVcc\ measured at the Tevatron\footnote{
The \LepII\ results for the W mass are not used for the TGC measurement, 
since they have been obtained under the assumption that W pairs are 
produced according to the \SM, whereas W production at 
the Tevatron does not involve the triple gauge vertex.}~\cite{MWPDG}.

In the analyses described below, a number of Monte Carlo models are
used to provide estimates of efficiencies and backgrounds as well as
the expected W-pair production and decay angular distributions for different 
TGC values. The majority of the Monte Carlo samples were
generated at \roots\ = 183~GeV with $\Mw=80.33$~\GeVcc.  All Monte
Carlo samples mentioned below, unless referred to as ``generator level'',
were passed through the full OPAL simulation program~\cite{GOPAL} 
and then subjected to the same reconstruction procedure as the data.

A \MC\ sample of signal events for the \xse\ analysis was produced
by the \Koralw~\cite{KORALW} generator with \WW\
production diagrams (class\footnote{
In this paper, the W pair production diagrams, {\em i.e.} 
$t$-channel $\nu_{\mathrm{e}}$ exchange and $s$-channel \Zgamma\ exchange, 
are referred to as ``CC03'', following the notation of \cite{LEP2YR}.} 
CC03) according to the \SM. The \Koralw\ generator has the most accurate 
simulation of ISR and a complete treatment of tau polarisation in \Qqtn\ 
events. This sample is used to estimate the W-pair selection efficiency. 
Other \MC\ samples used for systematic checks were generated with the   
same set of diagrams using the \Excalibur~\cite{EXCALIBUR} and 
\Grace~\cite{GRC4F} programs. To estimate the hadronization systematics,
\MC\ samples were produced by the \Pythia~\cite{PYTHIA} and
\Herwig~\cite{HERWIG} generators.

The \tgc\ studies rely mainly on the \Excalibur\ program which has been 
used to generate samples with anomalous \tgc s in order to calculate
the \tgc\ dependence of the selection efficiency and to obtain the
expected angular distributions used in the \tgc\ analysis.  

The separation between signal and background processes is complicated 
by the interference between the CC03 set of W-pair production diagrams 
and other four-fermion graphs, such as \Wenu, \Zee\ and \ZZ. To study 
the influence of interference effects in the four-fermion final states, 
the \Grace\ and \Excalibur\ Monte Carlo generators are used. In
both cases samples were generated using the full set of interfering 
four-fermion diagrams. These four-fermion samples were compared to samples
obtained with the same generator using only the CC03 diagrams.
Four-fermion samples were generated also with anomalous \tgc s.

Other background sources do not interfere with the signal.
The main one, \ZGqq, including higher order QCD diagrams, is simulated 
using \Pythia, with \Herwig\ used as an alternative to study possible 
systematic effects. Other background processes involving two fermions 
in the final state are studied using \Koralz~\cite{KORALZ} for \eemumu, 
\eetautau\ and \eenunu, and \Bhwide\cite{BHWIDE} for \eeee. 
Backgrounds from two-photon processes are evaluated using \Pythia, 
\Herwig, \Phojet~\cite{PHOJET} and the Vermaseren 
generator\cite{VERMASEREN}. It is assumed that the present centre-of-mass
energy of 183~\GeV\ is below the threshold for Higgs boson production.


\section{Event selection}
\label{sec:evsel} 
W-pair events decay into
fully leptonic (\WWlnln), semi-leptonic (\WWqqln) and fully hadronic 
(\WWqqqq) final states with expected branching fractions of 10.5\%, 43.9\% 
and 45.6\%, respectively. The \WW\ event selection consists 
of three distinct parts to identify each of these topologies. 
The selections used at 183 \GeV\ are 
based upon those used in the OPAL analysis at 
\roots~$\approx 172$~\GeV~\cite{mass172-analysis}.
The three selections are exclusive. 
Only events failing the  \WWlnln\ selection are considered as possible
\WWqqln\ candidates and only events failing both the \WWlnln\ and \WWqqln\
selections are considered as possible \WWqqqq\ candidates. 
The fully leptonic and semi-leptonic selections 
are separated into the individual lepton types to test
charged current lepton universality.

The detailed selection algorithms are described below. The efficiency
values and accepted background cross sections are summarised in 
Tables~\ref{tab:eff_matrix} and~\ref{tab:backs}. 

\renewcommand{\arraystretch}{1.2}
\begin{table}[htbp]
 \begin{center}
 \begin{tabular}{|l|cccccc|ccc|c|} \hline
 Event    &     \multicolumn{10}{c|}{Efficiencies[\%] for $\WW\rightarrow$}    \\    
 selection&\senen$\!\!$&\senmn$\!\!$&\sentn$\!\!$&\smnmn$\!\!$&\smntn$\!\!$&
    \stntn$\!\!$&\Qqen$\!\!\!$&\Qqmn$\!\!\!$& \Qqtn$\!\!\!$&\Qqqq$\!\!\!\!$\\ \hline
 \senen    &67.3 &  0.2 &  6.5 &  0.0 &  0.0 &  0.6 &  0.0 &  0.0 &  0.0 &  0.0 \\
 \senmn    & 2.2 & 69.9 &  5.5 &  0.5 &  5.8 &  1.2 &  0.0 &  0.0 &  0.0 &  0.0 \\
 \sentn    &12.2 &  8.7 & 58.2 &  0.0 &  1.6 &  8.9 &  0.0 &  0.0 &  0.0 &  0.0 \\
 \smnmn    & 0.0 &  1.2 &  0.1 & 73.7 &  5.7 &  0.6 &  0.0 &  0.0 &  0.0 &  0.0 \\
 \smntn    & 0.1 &  3.5 &  0.6 & 10.8 & 56.4 &  6.8 &  0.0 &  0.0 &  0.0 &  0.0 \\
 \stntn    & 0.7 &  0.9 &  4.9 &  1.3 &  6.2 & 44.1 &  0.0 &  0.0 &  0.0 &  0.0 \\ \hline
 \Qqen     & 0.0 &  0.0 &  0.1 &  0.0 &  0.0 &  0.1 & 84.0 &  0.1 &  5.8 &  0.1 \\ 
 \Qqmn     & 0.0 &  0.0 &  0.0 &  0.0 &  0.1 &  0.0 &  0.1 & 86.2 &  4.5 &  0.2 \\
 \Qqtn     & 0.0 &  0.0 &  0.0 &  0.0 &  0.0 &  0.1 &  4.5 &  4.5 & 64.6 &  0.6 \\ \hline
 \Qqqq     & 0.0 &  0.0 &  0.0 &  0.0 &  0.0 &  0.0 &  0.0 &  0.0 &  0.5 & 84.6 \\ \hline
 \end{tabular}
\end{center}
\caption{ Efficiency matrix for the 183~\GeV\ event selections
          determined using \Koralw\ (CC03) 
          Monte Carlo events. Each entry represents the percentage
          of generated  events for each decay channel  which are accepted 
          by the different selections.}
\label{tab:eff_matrix}
\end{table}
\renewcommand{\arraystretch}{1.0}

\begin{table}[htbp]
 \begin{center}
 \begin{tabular}{|l|r@{$\ \pm \ $}l|r@{$\ \pm \ $}l|
                    r@{$\ \pm \ $}l|r@{$\ \pm \ $}l|r@{$\ \pm \ $}l|} \hline
 \multicolumn{1}{|l|}{} &
 \multicolumn{10}{c|}{Accepted background \xses\ (fb)} \\
 \multicolumn{1}{|l|}{} &
 \multicolumn{10}{c|}{Event Selection $\WW\rightarrow$ } \\
 \multicolumn{1}{|l|}{Source} & 
 \multicolumn{2}{c|}{\Lnln} & 
 \multicolumn{2}{c|}{\Qqen} & \multicolumn{2}{c|}{\Qqmn}& 
 \multicolumn{2}{c|}{\Qqtn} &
 \multicolumn{2}{c|}{\Qqqq} \\ \hline 
 \Qqqq     &  0&0  &   2&2    &  1&2  &   6&4  &  220&50  \\
 \Qqen     &  0&0  &  56&28   &  1&1  &  53&11 &    0&0   \\
 \Qqll     &  0&0  &  65&19   & 29&4  &  74&7  &   72&6   \\
 \Qqnn     &  0&0  &   0&0    &  0&0  &  12&3  &    0&0   \\
 \Llnn     & 40&23 &   0&0    &  0&0  &   0&0  &    0&0   \\
 \Llll     &  9&2  &   1&1    &  0&0  &   1&1  &    0&0   \\
 \ZGqq     &  0&0  &  61&17   & 28&6  & 183&22 & 1370&150 \\
 \ZGll     &  6&1  &   2&1    &  1&1  &   6&2  &    0&0   \\
 Two-photon& 30&30 &  13&13   &  0&0  &   5&5  &    6&6   \\ \hline
 Combined  & 85&38 & 200&40   & 60&8  & 340&27 & 1670&160 \\ \hline
 \end{tabular}
\end{center}
\caption{Background \xses\ for the 183~\GeV\ \WW\ selections. 
  The values for the \Qqqq, \Qqen\ and \Lnln\
  sources are obtained from the difference between the full four-fermion
  and the corresponding CC03 \xses.   
  All errors include both statistical and systematic contributions. }
\label{tab:backs}
\end{table}

The first six background sources listed in Table~\ref{tab:backs} are
different four-fermion final states. Their expected \xses\ 
include both contributions from non-CC03 diagrams and the effects of 
interference. The four-fermion background \xses\
for each final state are calculated from the difference
between the accepted four-fermion \xse\ including all diagrams,  
and the accepted CC03 \xse. For this determination
the \Grace\ Monte Carlo generator is used. The
associated systematic uncertainty is estimated by comparing the
predictions of \Grace\ and \Excalibur. 
At the current level of statistical precision such interference 
effects are small within our experimental angular acceptance.


\subsection{\WWlnln\ event selection}
\label{subs:lnlnsel}

 Fully leptonic \WW\ events are identified as an acoplanar pair
 of charged leptons with missing momentum. The event selection is 
 unchanged from that used in reference \cite{mass172-analysis} 
 (described in detail in \cite{acopll-analysis}). 
 In the $\sqrt{s}\approx 183$~\GeV\ data sample, 78 events are 
 selected as \WWlnln\ candidates. 
 
 The selection efficiency is estimated to be $(78.0\pm2.3)\%$
 where the error includes systematic uncertainties.
 The efficiencies for the individual channels
  are given in Table~\ref{tab:eff_matrix}.
 The dominant systematic uncertainty on the selection efficiency arises 
 from differences observed when comparing Monte Carlo samples 
 which have different implementations of both initial and final state 
 radiation effects and the modelling of tau decays. In addition, 
 systematic errors were included to account for data/Monte Carlo 
 disagreement (0.8\%), the possibility of events 
 being rejected due to off-momentum beam particles
 which can deposit significant energy in the forward detectors (0.5\%), 
 and the knowledge of the trigger efficiency (0.4\%). 
 The expected background \xses\ from
 Standard Model processes are given in Table~\ref{tab:backs}. 


\subsection{\WWqqln\ event selection}
 \label{subs:qqlnsel}

The \WWqqln\ selection consists of three separate selections, one
for each lepton flavour. The \WWqqtn\ selection
is applied only to events which fail both the \WWqqen\ and
\WWqqmn\ selections.  

The \WWqqln\ event selection for the 183~\GeV\ data is a modified 
version of the 172~\GeV\ selection described in detail in 
\cite{mass172-analysis}.
A looser set of preselection cuts is used since the lepton energy spectrum 
is broader at 183~\GeV\ due to the increased boost.
The set of variables used in the likelihood selections has been also 
modified. In the \WWqqtn\ sample there is a significant background from
hadronic decays of single-W four-fermion events (\Wenu) 
and an additional likelihood 
selection is used to reduce this. This selection is only applied to
\WWqqtn\ events where the tau is identified as decaying in the 
single-prong hadronic channel. Finally, in order to reduce the \ZZ\ 
background, events passing the 
\WWqqen\ likelihood selection are rejected if there is evidence for
a second energetic electron. A similar procedure is applied to the
\WWqqmn\ selection.

In total, 361 events are identified as \WWqqln\ candidates, 
of which 140 are selected as 
\WWqqen, 120 as \WWqqmn\ and 101 as \WWqqtn. 
The efficiencies of the \WWqqln\ selection for the individual 
channels are given in Table~\ref{tab:eff_matrix}.
The efficiencies include corrections ($\approx 0.5\%$) which account 
for observed differences between the data 
and the Monte Carlo simulation as described in \cite{mass172-analysis}. 
The efficiencies are also corrected by a factor of $0.991\pm0.003$ for the
$\WWqqen$ and $\WWqqmn$ selections and by a factor of $0.985\pm0.005$
for the $\WWqqtn$ selection to  
account for events being rejected due to the presence of off-momentum beam 
particles which can deposit significant energy in the forward detectors. 
A systematic error is assigned for possible 
tracking losses which are not modelled by
the Monte Carlo simulation of the OPAL detector. This uncertainty is evaluated 
using selections designed to identify \WWqqen\ and \WWqqmn\ events 
where the track was either not reconstructed or was poorly measured. 
Possible biases due to fragmentation uncertainties are studied by comparing
fully simulated Monte Carlo \WWqqln\ samples with fragmentation performed
using \Pythia\ and \Herwig. Other systematics are estimated using samples 
generated with different
Monte Carlo models and different input W boson masses and beam energies. 
Table~\ref{tab:qqln-sys} lists the sources of the 
systematic uncertainties on the  selection efficiencies.

\begin{table}[htbp]
\begin{center}
\begin{tabular}{|ll|c|c|c|c|} \hline
 \multicolumn{2}{|l|}{} &
 \multicolumn{4}{c|}{Signal efficiency  error (\%) }    \\
 \multicolumn{2}{|l|}{} &
 \multicolumn{4}{c|}{Event selection $\WW\rightarrow$ } \\
&Source of uncertainty            &\Qqen\ &\Qqmn\ & \Qqtn\  & \Qqqq\ \\ \hline
a) & Statistical                      & 0.3 & 0.3 & 0.4 & 0.2 \\ 
b) & Comparison of MC models          & 0.6 & 0.4 & 0.5 & 0.3 \\
c) & Data/Monte Carlo                 & 0.6 & 0.5 & 0.6 & 0.6 \\ 
d) & Tracking losses                  & 0.4 & 0.4 & 0.4 & $-$ \\
e) & Detector occupancy               & 0.3 & 0.3 & 0.3 & $-$ \\
f) & Fragmentation                    & 0.2 & 0.1 & 0.6 & 0.8 \\
g) & \Mw\ dependence ($\pm0.09$~\GeVcc) & 0.1 & 0.2 & 0.1 & 0.1 \\ 
h) & Beam energy dependence           & 0.1 & 0.1 & 0.1 & 0.1 \\ \hline
& Combined                            & 1.1 & 0.9 & 1.2 & 1.0 \\ \hline
 \end{tabular}
 \end{center}
 \caption{ Sources of uncertainty on the \WWqqln\ and \WWqqqq\ 
           selection efficiencies.}
 \label{tab:qqln-sys}
\end{table}

Table~\ref{tab:backs} shows the corrected background \xses\ and their
total uncertainties for the three \Qqln\ selections. The systematic 
errors on the
expected background \xses\ are dominated by differences between data
and Monte Carlo for the two-fermion backgrounds and by differences between
generators in the case of the four-fermion backgrounds. The systematic errors
on the four-fermion backgrounds were estimated by comparing the expectations 
of \Grace, \Excalibur\ and \Pythia. 

An initial estimate of the \ZGqq\ background is obtained from Monte Carlo. 
This estimate is then corrected using data to account
for possible uncertainties in the Monte Carlo modelling, {\em e.g.} the 
lepton fake rate in \ZGqq\ events. Several
methods are used to estimate the correction factors to 
the Monte Carlo estimates of the \ZGqq\ background. 
The central value is obtained using `fake' \ZGqq\ events formed
by boosting \Zqq\ events from at $\sqrt{s}=91.2$~\GeV\ along the $z$-axis 
according to the invariant mass distribution of the hadronic system in
\ZGqq\ events at \roots=183~\GeV~\cite{fpair183-analysis}. 
This procedure is applied to both
data and Monte Carlo. The correction factor is
determined from the relative fractions (data/Monte Carlo) 
of `fake' events which pass the \WWqqln\ selections.
As a result, the Monte Carlo estimates of the \ZGqq\ background 
are scaled by a factor of $1.3\pm0.3$ for the 
\WWqqen\ selection, $1.0\pm0.1$ for the \WWqqmn\ selection and 
by $1.1\pm0.1$ for the \WWqqtn\ selection. The errors
reflect the spread of values obtained using alternative methods, 
{\em e.g.} using events passing the preselection and then fitting 
Monte Carlo signal (\WWqqln) and background (\ZGqq) components 
to the kinematic distributions of the data.

In the \WWqqtn\ selection there is a non-negligible background
from hadronic decays of single-W events (\Wenu ) where the electron goes
undetected down the beam pipe. This background is determined by the number
of fake tau candidates formed out of the fragmentation products in the
hadronic decay of the W. A correction factor to the Monte Carlo background
is obtained using fake single-W events formed by removing
the lepton candidate from selected \WWqqen\ and \WWqqmn\ events (both
in data and Monte Carlo). The ratio of \WWqqtn\ selection efficiencies
(data/Monte Carlo) for these fake events, $0.8\pm0.1$, is used to scale the 
Monte Carlo estimate of the \Wenu\ background.


\subsection{\WWqqqq\ event selection}
\label{subs:qqqqsel}

The selection of fully hadronic \WWqqqq\ events is performed in two 
stages using a cut-based preselection followed by a likelihood selection 
procedure. The general features of this selection are similar to those 
used previously at 172 \GeV~\cite{mass172-analysis}, although it
has been re-optimised to improve the rejection of the 
dominant background arising from hadronic \Zqq\ decays.

All events which are classified as hadronic events~\cite{TKMH} 
and which have not been selected by either the \WWlnln\ or \WWqqln\ selections 
are considered as candidates for the \WWqqqq\ selection.
Tracks and calorimeter clusters are combined into four jets using the
Durham $k_\perp$ algorithm~\cite{DURHAM} and the total momentum and energy of 
each of the jets are corrected for double-counting of energy~\cite{GCE}.
To remove events which are clearly inconsistent with a hadronic \WWqqqq\
decay, candidate events are required to satisfy the following 
preselection criteria:

\begin{itemize}
       \item  The fitted invariant mass of the event scaled by the 
              collision energy, $\sqrt{s'/s}$, must be greater than 0.75.
       \item  The visible energy of the event must be greater than 
              $0.7 \sqrt{s}$.
       \item  The energy of the most energetic electromagnetic cluster must 
              be less than $0.3 \sqrt{s}$.
       \item  Each jet is required to contain at least one  
              charged track.
       \item  The quantity\footnote{With the jets ordered by energy, 
              the quantity $j_{\rm ang}$ is defined as 
              \mbox{$\frac{E_{4}}{\sqrt{s}} (1-c_{12} c_{13} c_{23})$} where 
              $c_{ij} = \cos\theta_{ij}$, $\theta_{ij}$ being the angle 
              between jets $i$ and $j$,
              and $E_{4}$ is the energy of the fourth (lowest energy) jet.
              This quantity is strongly peaked towards zero for the dominant
              \ZGqq\ background process.} $j_{\rm ang}$ must be greater than 
              0.05.
       \item  The logarithm of the QCD matrix element
               $W_{420}$~\cite{QCDMATRIX}, calculated using the jet 
               momenta as estimates of the parton momenta, is required
               to be less than zero. $W_{420}$ is an event weight formed from
               the tree level $O(\alpha_s^2)$ matrix element~\cite{ERT}
               for the QCD process ($\epem\rightarrow\qqqq,\qqgg$). It is 
               assigned the largest value of any permutation of 
               associated jets to partons.
\end{itemize}

This preselection rejects $97.7$\% of the \Zqq\ events
which comprise the dominant background in the \WWqqqq\ channel.
The preselection efficiency for the hadronic \WWqqqq\ decays 
is estimated to be $88.8$\%. In total, 524 candidates pass the preselection.

Events satisfying the preselection cuts are classified as signal or 
background based upon a four variable likelihood selection. The 
following likelihood variables are used since they provide a good separation
between the hadronic \WWqqqq\ signal and the dominant \Zqq\ background 
process for a minimum number of variables used:
\begin{itemize}
        \item  $\log(y_{45})$, the logarithm of the value of the Durham 
               jet resolution 
               parameter at which an event is reclassified from four jets  
               to five jets.
        \item  $\log(W_{420})$, the logarithm of the QCD matrix 
               element~\cite{QCDMATRIX}.
        \item  $|\cos\theta_{\rm N-R}|$, the cosine of the 
               modified Nachtmann-Reiter angle\footnote{
               With the jets ordered by energy, the
               quantity $\cos\theta_{\rm N-R}$ is defined
               as $\frac{(\vec{p}_1-\vec{p}_2)\cdot(\vec{p}_3-\vec{p}_4)}
               {|\vec{p}_1-\vec{p}_2||\vec{p}_3-\vec{p}_4|}$.
               This variable, which is sensitive to correlations between the 
               underlying parton momenta, tends to be flat for the dominant 
               \ZGqq\ background and somewhat peaked for the \WWqqqq\ signal 
               events.}~\cite{COSNR}.
        \item  the sphericity of the event.
\end{itemize}

Rather than using the product of the individual probability density 
functions to construct a classic likelihood discriminator, a coordinate 
transformation technique has been developed to reduce correlations 
between the four input variables~\cite{Karlen}.

\begin{figure}[htbp]
  \begin{center}
    \leavevmode
    \epsfig{file=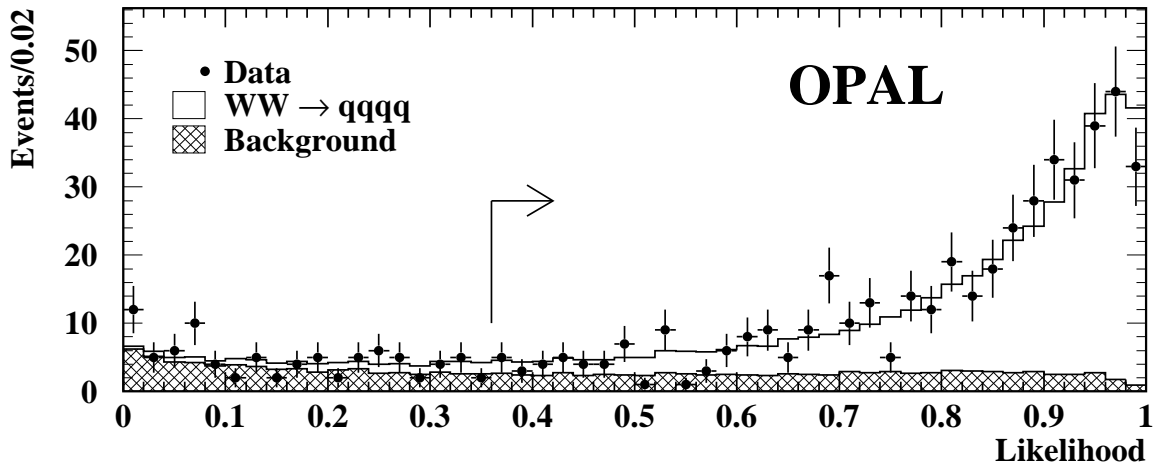,width=\linewidth}
    \caption{\sl
      The distribution of the likelihood discriminant used to select 
      events in the \WWqqqq\ selection is shown for all preselected events.
      The points indicate the data and the histogram represents the Monte
      Carlo expectation where the hatched area
      shows the estimated contribution of the total background.
      The selection cut is indicated by the arrow.}
    \label{fig:qqqqlike}
  \end{center}
\end{figure}

An event is then selected as a hadronic \WWqqqq\ decay if the likelihood
discriminant variable $L$, shown for data and Monte Carlo in Figure
\ref{fig:qqqqlike}, is greater than 0.36. This cut value was chosen to 
maximise the product of signal purity and efficiency.

In the OPAL 172~\GeV\ \xse\ measurement\cite{mass172-analysis}, 
a 5\% improvement in the statistical error was achieved by counting the 
event weights of all preselected events rather than simply specifying a 
cut on the discriminant variable. In this analysis, the use of event weights 
improves the statistical error on the \xse\ measurement by less 
than 2\% and results in an increased systematic uncertainty.
For this reason, the more straightforward counting method is used.

The overall efficiency of the \WWqqqq\ event selection is estimated
from the \Koralw\ Monte Carlo simulation to be $(84.6 \pm 1.0)\%$ where the 
error is an estimate of all known systematic uncertainties;
the individual components are listed in Table~\ref{tab:qqln-sys}.
The total expected background \xse\ from non-CC03 diagrams
is estimated to be $(1.67 \pm 0.16)$ pb where the error represents 
the systematic uncertainty. The processes contributing to 
the background \xse\ are listed in Table \ref{tab:backs}.
In total, 438 candidates pass the selection.

The main source of uncertainty for both the signal efficiency and the
background \xse\ is related to the modelling of the fragmentation process.
This uncertainty is estimated by comparing the selection
efficiency for both signal and background events using an alternative QCD 
Monte Carlo model (\Herwig).
In addition, the parameters $\sigma_q$, $b$, $\Lambda_{QCD}$, and $Q_{0}$ 
of the \Jetset\ fragmentation model are varied by one standard 
deviation about their tuned values~\cite{qqqqQCD}. 
The Monte Carlo modelling of the data is further studied by comparing
the distributions of the four likelihood variables seen in the data with
various Monte Carlo samples.
The signal efficiency evaluated using \Koralw\ is compared to alternate
generators (\Excalibur, \Pythia\ and \Grace) to test the Monte
Carlo description of the underlying hard process.
Uncertainties related to the beam energy and W mass are evaluated with 
\Pythia\ samples generated over a range of values.
In each case, the observed differences are taken as an estimate of the
systematic uncertainty.


\section{\WW\ \xse\ and W decay branching fractions}
 \label{sec:xsect}
\subsection{Cross-section and branching fraction results}
 \label{subs:xseres}

 The observed numbers of selected \WW\ events are used to measure the
 \WW\ production \xse\ and the W decay branching fractions 
 to leptons and hadrons.  The measured \xse\ corresponds to that for 
 W pair production via the CC03 diagrams. 

 \begin{table}[htbp]
 \begin{center}
 \begin{tabular}{|l|r|r|r|c|} \hline
  Selected as & Efficiency [\%] & Purity [\%] & Expected & Observed \\ \hline
  \WWlnln  & $78.0 \pm 2.3 $ & $93.8$ & $ 78.9\pm 2.3$ &  78 \\ 
  \WWqqln  & $84.7 \pm 1.3 $ & $90.2$ & $371.2\pm 8.2$ & 361 \\
  \WWqqqq  & $84.6 \pm 1.0 $ & $78.3$ & $442.6\pm12.4$ & 438 \\ \hline
  Combined & $84.4 \pm 0.8 $ & $85.2$ & $892.7\pm19.2$ & 877 \\ \hline
 \end{tabular}
 \end{center}
\caption{Observed numbers of candidate events in each \WW\ decay
 topology for an integrated luminosity of 
 ($\intLdtfull\pm\dLtot$)~\Ipb\ at ($\rroots\pm\droots$)~\GeV,
  together with expected numbers of events, assuming 
  $\Mw=(80.40\pm0.09)$~\GeVcc\ and \SM\ branching fractions. The expected 
  numbers of events include systematic uncertainties from the efficiency,
  luminosity, beam energy, \WW\ \xse\ (2\%)
  and \Mw. In this table cross contamination between \WW\ events from different
  topologies has been included as background ({\em i.e.} excluded from the
  efficiency numbers).  The errors on the combined numbers account for 
  correlations.}
\label{tab:sel_sum}
\end{table}

 Table~\ref{tab:sel_sum} summarises the event selections in the
 three \WW\ decay topologies. The efficiencies refer to CC03 \WW\ events.
 The expected number of events is calculated using
 the \Gentle\ \xse\ of \GENTxs~pb assuming a W mass of
 $80.40$~\GeVcc~\cite{MWPDG} and a centre-of-mass energy
 of $\rroots$~\GeV. The systematic uncertainties on the expected numbers of
 events include contributions from the current errors of
 $\pm$0.09~\GeVcc\ on \Mw\ and $\pm$0.05~\GeV\ on the \Com\ energy
 (both below 0.1\%) and a 2\% theoretical uncertainty on the \xse\
 calculation. The data are consistent with the Monte Carlo expectation.

 The selected numbers of events can be used to determine the \WW\ 
 CC03 production \xses\ separately into each channel: 
\begin{eqnarray*}
 \sigma(\WWlnln) & = & (1.64 \pm 0.20 \pm 0.07) \; {\mathrm{pb}}, \\
 \sigma(\WWqqln) & = & (6.68 \pm 0.39 \pm 0.12) \; {\mathrm{pb}}, \\
 \sigma(\WWqqqq) & = & (7.07 \pm 0.43 \pm 0.21) \; {\mathrm{pb}}. 
\end{eqnarray*}

As in \cite{mass172-analysis}, the \WW\ \xse\ and branching 
fractions are 
measured using the observed events from the ten separate channels
summarised in Table~\ref{tab:selected_channels}. 
Three different maximum likelihood fits are performed. In the first case,
\sigccthree(183~\GeV), Br(\Wev), Br(\Wmv) and Br(\Wtv) are
extracted under the assumption that
\begin{eqnarray*}
     {\rm Br}(\Wev)+{\rm Br}(\Wmv)+{\rm Br}(\Wtv)+{\rm Br}(\Wqqp) & = & 1.
\end{eqnarray*}
In the second fit, charged current lepton universality is imposed and in 
the third fit, the \WW\ \xse\ is determined assuming Standard Model branching
fractions. The results are summarised in Table~\ref{tab:xsecbr_results}.

\begin{table}[htbp]
 \begin{center}
 \begin{tabular}{|l|r|r|r|c|} \hline
  Selected as & Expected signal & Expected back. & Total & Observed \\ \hline
  \WWenen  & $  8.6 \pm 0.4 $ & $  0.2\pm0.4 $  & $  8.8 \pm 0.6$ & 12  \\ 
  \WWenmn  & $ 17.5 \pm 0.6 $ & $  0.2\pm0.1 $  & $ 17.8 \pm 0.6$ & 11  \\
  \WWentn  & $ 16.7 \pm 0.6 $ & $  1.0\pm1.0 $  & $ 17.6 \pm 1.6$ & 20  \\ 
  \WWmnmn  & $  9.3 \pm 0.3 $ & $  1.4\pm1.0 $  & $ 10.7 \pm 1.0$ & 13  \\ 
  \WWmntn  & $ 14.6 \pm 0.6 $ & $  1.1\pm1.0 $  & $ 15.7 \pm 1.2$ & 15  \\
  \WWtntn  & $  7.4 \pm 0.4 $ & $  0.8\pm0.6 $  & $  8.2 \pm 0.7$ &  7  \\ \hline
  \WWqqen  & $118.4 \pm 2.8 $ & $ 11.3\pm2.3 $  & $129.8 \pm 3.6$ & 140 \\
  \WWqqmn  & $119.8 \pm 2.7 $ & $  3.4\pm0.5 $  & $123.2 \pm 2.8$ & 120 \\
  \WWqqtn  & $ 99.1 \pm 3.0 $ & $ 19.2\pm1.6 $  & $118.2 \pm 3.3$ & 101 \\ \hline
  \WWqqqq  & $347.1 \pm 8.3 $ & $ 95.5\pm9.2 $  & $442.6 \pm12.4$ & 438 \\ \hline
  Combined & $758.4 \pm16.5 $ & $134.3\pm9.6 $  & $892.7 \pm19.2$ & 877 \\ \hline
 \end{tabular}
 \end{center}
\caption{Observed numbers of candidate events in each \WW\ decay
  channel for an integrated luminosity of 
 ($\intLdtfull\pm\dLtot$)~\Ipb\ at ($\rroots\pm\droots$)~\GeV\
  together with expected numbers of signal and background events,
  assuming \Mw=80.40 $\pm$ 0.09~\GeVcc. The predicted numbers of
  signal events include systematic uncertainties from the efficiency,
  luminosity, beam energy, \WW\ \xse\ 
  and \Mw, while the background
  estimates include selection and luminosity uncertainties. The errors
  on the combined numbers account for correlations. }
\label{tab:selected_channels}
\end{table}

\renewcommand{\arraystretch}{1.1}
\begin{table}[htbp]
 \begin{center}
 \begin{tabular}{|clc|l|l|l|} \hline
& Fitted parameter&   &  \multicolumn{3}{c|}{Fit assumptions : }    \\   
& 183 \GeV\ data &   &   & Lepton universality & SM branching fractions\\ \hline
& Br(\Wev) &  & $ 0.121\pm.010\pm.003$ & & \\ 
& Br(\Wmv) &  & $ 0.107\pm.009\pm.003$ & & \\ 
& Br(\Wtv) &  & $ 0.094\pm.011\pm.003$ & & \\ 
& Br(\Wlv) &  &     &  $0.108\pm.004\pm.002$ &  \\ 
& Br(\Wqqp) &  & $0.678\pm.013\pm .005$ & 
                              $0.676\pm.013\pm.005$ &  \\ \hline 
& \sigccthree(183~\GeV) [pb]& & $15.33 \pm 0.61 \pm 0.27$ 
                              & $15.43 \pm 0.61 \pm 0.26$ 
                              & $15.43 \pm 0.61 \pm 0.26$\\ \hline \hline
& Fitted parameter &  &  \multicolumn{3}{c|}{Fit assumptions : }    \\   
& 161-183 \GeV\ data &  &   & \multicolumn{2}{l|}{Lepton universality} \\ \hline
& Br(\Wev) &  & $ 0.117\pm.009\pm.002$ & \multicolumn{2}{l|}{} \\ 
& Br(\Wmv) &  & $ 0.102\pm.008\pm.002$ & \multicolumn{2}{l|}{} \\ 
& Br(\Wtv) &  & $ 0.101\pm.010\pm.003$ & \multicolumn{2}{l|}{} \\ 
& Br(\Wlv) &  &  &  \multicolumn{2}{l|}{$0.107\pm.004\pm.002$} \\ 
& Br(\Wqqp) &  & $0.680\pm.012\pm .005$ & 
           \multicolumn{2}{l|} {$0.679\pm.012\pm.005$} \\ \hline 
 \end{tabular}
\end{center}
\caption{ Summary of \xse\ and branching fraction results from 
          the 183~\GeV\ data and the branching fraction results 
          from the combination of the 161~\GeV, 172~\GeV\
          and 183~\GeV\ data. The results from three 
          different fits described in the text are shown. 
          The correlations between the branching fraction measurements
          from the fits without the assumption of lepton universality 
          are less than 27\%.}
\label{tab:xsecbr_results}
\end{table}
\renewcommand{\arraystretch}{1.0}

The hadronic branching fraction can be interpreted as a measurement of  
the sum of the squares of the six elements of the CKM
mixing matrix, \Vij, which do not involve the top quark \cite{LEP2YR}:
\begin{eqnarray*}
     {{\Br(\Wqq)}\over{(1-\Br(\Wqq))}} & = & 
 \left( 1+\frac{\alpha_s(\Mw)}{\pi} \right)
\sum_{i={\mathrm{u,c}}; \, j={\mathrm{d,s,b}}} \Vij^2,
\end{eqnarray*}
where $\alpha_s(\Mw)$ is taken to be $0.120\pm0.005$.
The branching fraction $\Br(\Wqqp)$ from the 161 -- 183~\GeV\
data obtained from the fit assuming lepton universality yields,
\begin{eqnarray*}
 \sum_{i={\mathrm{u,c}};\, j={\mathrm{d,s,b}}}
  \Vij^2 & = & 2.04\pm0.11\pm0.05.
\end{eqnarray*}
This is consistent with a value of 2 which is expected from unitarity.
Using the experimental knowledge\cite{pdg} of the sum, 
$\Vud^2+\Vus^2+\Vub^2+\Vcd^2+\Vcb^2 = 1.05\pm0.01$, the above result can
be interpreted as a measure of \Vcs, which is the least well determined 
of these elements: 
\begin{eqnarray*}
  \Vcs & = & 0.99 \pm 0.06 \pm 0.02.
\end{eqnarray*}

\begin{figure}[tbhp]
 \epsfxsize=\textwidth
 \epsffile{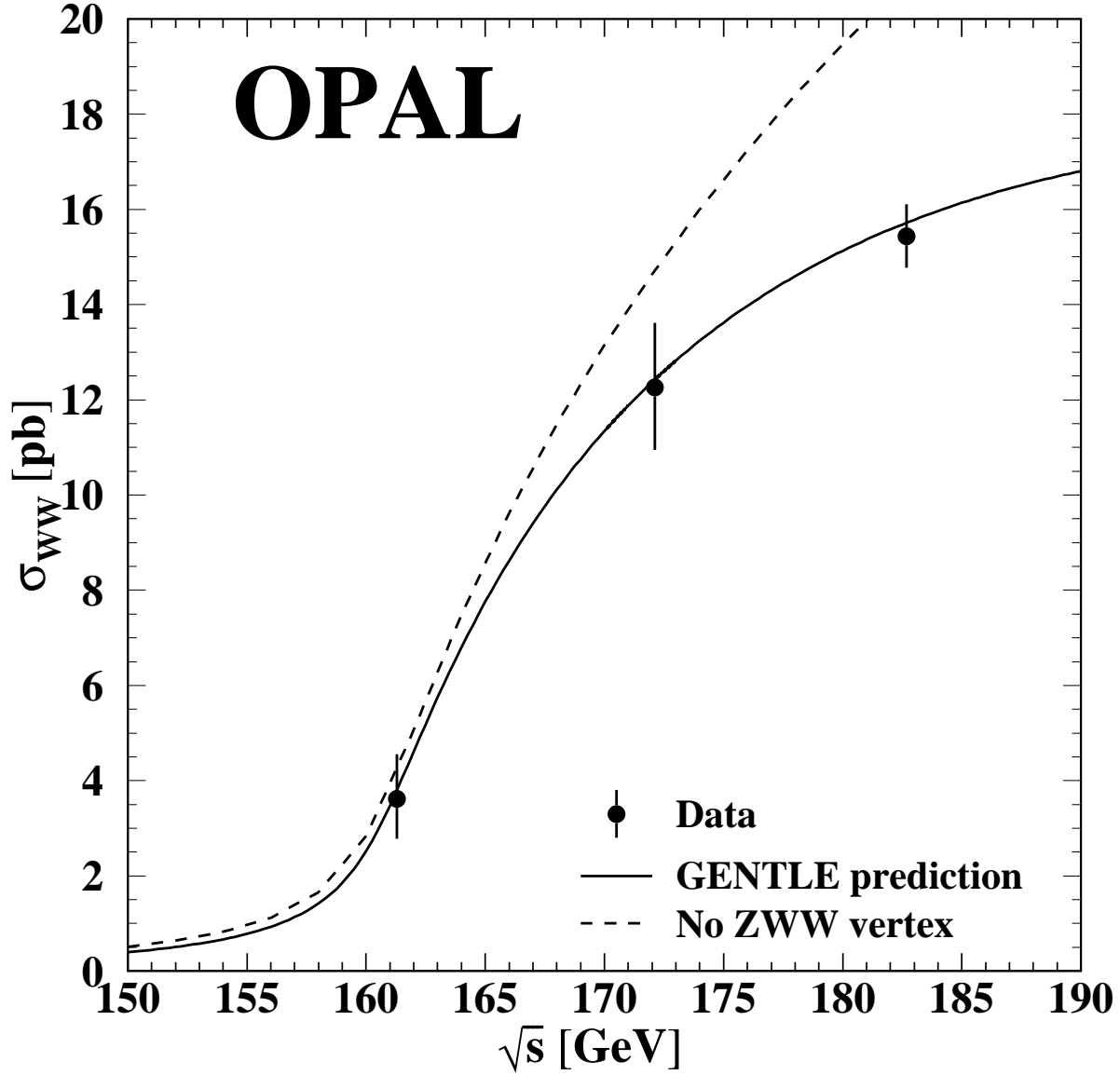}
   \caption{\sl
     The dependence of \sigccthree\ on
     $\protect\roots$, as predicted by \Gentle\ for $\Mw = 80.40$~\GeVcc. 
     The
     \WW\ \xses\ measured at $\protect\roots=182.7$~\GeV\ (this
     work), at $\protect\roots=161.3$~\GeV~\protect\cite{opalmw161} and
     $\protect\roots=172.1$~\GeV~\protect\cite{mass172-analysis} are
     shown.  The error bars include statistical and systematic
     contributions. The dashed curve shows the expected \xse\ if the
     ZWW couplings are zero.}
 \label{fig:sigmaww} 
\end{figure}

The measured \WW\ production \xse\ at $\roots=\rroots$~\GeV\ is
shown in Figure~\ref{fig:sigmaww} together with the recent OPAL
measurements of \sigccthree\ at $\roots=161.3$~\GeV~\cite{opalmw161} and
at $\roots=172.1$~\GeV~\cite{mass172-analysis}. Figure~\ref{fig:sigmaww}
also shows the \Gentle\ prediction which is in excellent agreement with
the data. On the other hand, the \xse\ calculated without
the contribution of the WWZ vertex, corresponding to anomalous
couplings \dgz=--1 and \dkz=--1 (dashed line in Figure~\ref{fig:sigmaww}) 
fails to describe the data.


\subsection{TGC analysis using the \xse}
\label{subs:tgcrate}
A quantitative study of TGCs from the W-pair event yield is performed by 
comparing the numbers of observed events in each of the three event
selection channels with the expected number which is parametrised
as a second-order polynomial in the TGCs. This parametrisation is
based on the linear dependence of the triple gauge vertex Lagrangian
on the TGCs, corresponding to a second-order polynomial dependence of 
the \xse. The polynomial coefficients are calculated from
the expected \xse\ in the presence of anomalous 
couplings determined with \Gentle\ and the slight dependence of our 
selection efficiency on the TGCs is obtained from the \Excalibur\ 
Monte Carlo samples.  
The \SM\ values for the W branching fractions are used.
The background, which originates predominantly from \ZGqq\ events, 
is assumed to be independent of the \tgc s. 

The probability to observe the measured number of candidates, given the 
expected value, is calculated using a Poisson distribution.
The product of the three probability distributions 
corresponding to the three event selection channels is taken as the \xse\
likelihood function.

The following sources of systematic uncertainty on the expected
number of events are considered.
\begin{itemize}
\item The theoretical uncertainty in the expected \xse\
which is obtained by comparing the \xses\ obtained  
from the \Gentle\ and \Excalibur\ programs. This uncertainty depends
on the anomalous couplings and has a typical size of 2\%.
\item The small effect of the uncertainties in the W mass from the
Tevatron measurement, (80.40 $\pm$ 0.09)~\GeVcc~\cite{MWPDG}, and the \LepII\ 
\Com\ energy, ($\rroots\pm\droots$)~\GeV, on the total \xse\ 
(less than 0.1\% each).
\item The uncertainties in the selection efficiencies and accepted
background \xses, as listed in 
Tables~\ref{tab:backs} and \ref{tab:sel_sum}. 
\item The uncertainty in the luminosity, of 0.5\%.
\end{itemize}
The systematic uncertainties are incorporated into the TGC fit by
allowing the expected numbers of signal and background events to 
vary in the fit and constraining them to have a Gaussian distribution
around their expected values with their systematic errors taken as the
width of the distributions. The systematic uncertainties, excluding 
those on efficiency and background, are 
assumed to be correlated between the three different event selections.
 
Data from lower centre-of-mass energies are included assuming all
systematic errors to be fully correlated between energies. 
The corresponding \LL\ curves are used in combination with the
results of the angular distribution analyses which are described
in the following sections. The full set of results are then presented
in Figure~\ref{fig:rateshape}, where the cross section contributions
are shown as dotted lines. 


\section{Analysis of the \WWqqln\ angular distributions}
\label{sec:qqln}
The analysis of the \WWqqln\ channel is performed in
three different ways using optimal observables (OO), a binned
maximum likelihood (BML) fit, and the spin density matrix (SDM).
These three methods will be described in the following sub-sections. 
All methods use the same selection and reconstruction procedure.
The selection procedure starts from the event 
sample used for the total \xse\ analysis with further cuts 
imposed in order to assure a reliable reconstruction of the event 
kinematics and to reduce further the background.


\subsection{Event selection and reconstruction}
\label{subs:qqlnrec}
The selection of \Qqen\ and \Qqmn\ events for the total \xse\
analysis results in a single track 
being identified as the most likely lepton candidate. 
The electron momentum vector is reconstructed  
by the tracking detectors and the energy is measured in 
the electromagnetic calorimeters. In the case of muons, the momentum 
measured using the tracking detectors is used. The remaining tracks and 
calorimeter clusters in the event are grouped into two jets
using the Durham $k_\perp$ algorithm~\cite{DURHAM}. The total 
energy and momentum of each of the jets are calculated with the method
described in~\cite{GCE}. A one-constraint kinematic fit is then performed 
on the events, requiring energy-momentum conservation and allowing for
a massless neutrino. Events are accepted if this fit converges
with a probability larger than 0.001. This cut rejects about 2\% of
the signal events and 10\% of the background.   

To improve the resolution in the angular observables used for the 
\tgc\ analysis, we perform a further kinematic fit with three
constraints requiring energy-momentum conservation and the fitted masses
of both the hadronic and the leptonic systems to be constrained
to the average Tevatron W mass, \Mw=80.40 \GeVcc~\cite{MWPDG},
within the W width\footnote{
 The W mass distribution is treated as a Gaussian in the kinematic fit. 
 However, in 
 order to simulate the expected Breit-Wigner form of the W mass spectrum,
 the variance of the Gaussian is updated at each iteration of the 
 kinematic fit in such a way that the probabilities of observing 
 the current fitted W mass are equal whether calculated
 using the Gaussian distribution or using a Breit-Wigner.}.
We demand that the kinematic fit converges with a probability larger than 
0.001. For the $\approx$3\% of events which fail at this point we revert
to using the results of the fit without the W mass constraint.
  
The selection of the \Qqtn\ events results in the identification
of the most likely tau decay candidate
classified as an electron, muon, one-prong hadronic or three-prong
hadronic decay. The remaining tracks and calorimeter clusters in the 
event are grouped into two jets as described above. However,
these events cannot be reconstructed in the same way 
since there is more than one unobserved neutrino. 
Nevertheless, as the tau is highly relativistic, its
flight direction can be approximated by the
direction of its charged decay products, leaving us with four unknown 
quantities, the tau energy and the prompt neutrino three-momentum vector.
A one-constraint kinematic fit is performed on the event 
requiring energy-momentum conservation and equality between the masses
of the hadronic and leptonic systems. This fit is required to converge
with a probability larger than 0.001. In addition, the angle between the
two jets in the hadronic system, $\theta_{jj}$, is required to
satisfy $\cos\theta_{jj}<-0.2$, and the angle between the tau and
the closest jet is required to exceed $20^o$. These cuts reject
20\% of the signal and 42\% of the background. Furthermore, these
cuts suppress those \Qqtn\ events which are 
correctly identified as belonging to this decay channel 
but where the tau decay products are not identified correctly, leading to 
an incorrect estimate of the tau flight direction or its charge.
The fraction of such events in the \Qqtn\ sample
is reduced by these cuts from 13\% to 8.5\%.

After these cuts, 324 \Qqln\ candidates remain 
(135 \Qqen, 116 \Qqmn\ and 73 \Qqtn). The remaining background fraction,
assuming \SM\ \xses\ for signal and background processes,
is 6.2\%, not including cross-migration between the three lepton
channels. The background sources are: four-fermion processes after subtracting
the contribution of the CC03 diagrams (3.8\%), \ZGqq\ (1.9\%),
\WWqqqq\ and \Lnln\ (0.3\%) and two-photon reactions (0.2\%). 

In the reconstruction of the \Qqln\ events we obtain $\Cthw$ by summing 
the kinematically fitted four-momenta of the two jets. 
The decay angles of the leptonically decaying W are obtained from the 
charged lepton four-momentum, after boosting back to the parent W rest frame.
In the hadronically
decaying W we are left with a twofold ambiguity in assigning the jets to the
quark and antiquark. This ambiguity is taken into account in the analyses
described below. 

\begin{figure}[tbhp]
 \epsfxsize=\textwidth
 \epsffile{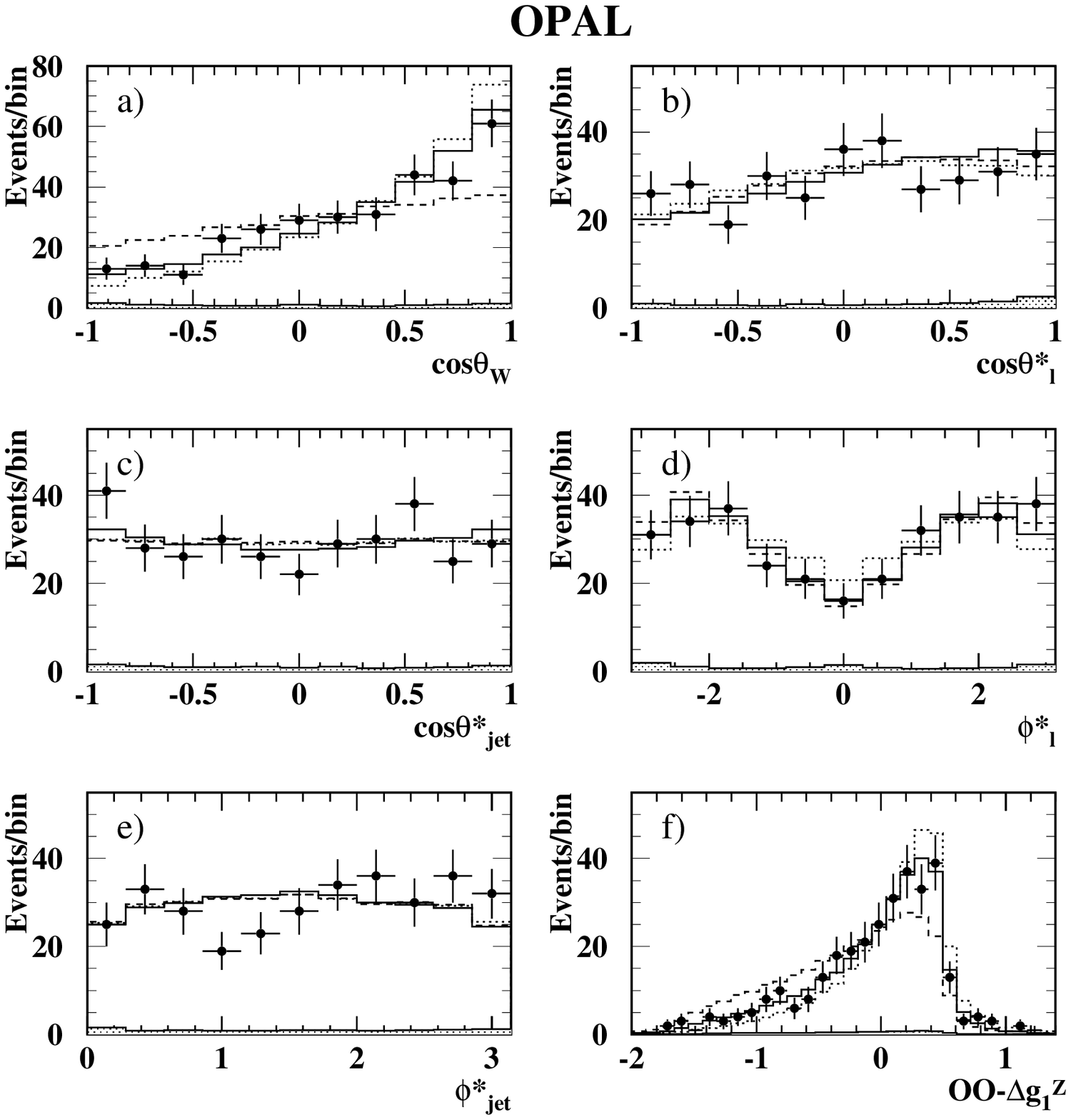}
   \caption{\sl
    Distributions of the kinematic variables \Cthw, \Cthstl,
    \Cthstj, \Phistl, \Phistj\ and the optimal observable corresponding
    to \dgz, as obtained from the \Qqln\ events. The solid points
    represent the data. The histograms show the expectation of the 
    \SM\ (solid line) and the cases of \dgz=+1 and --1 (dotted and dashed 
    lines respectively). The shaded histogram shows the 
    non-\Qqln\ background.
    Notes: 1. In the case of W$^+\ra\bar{\ell}\nu$ 
    decays the value of $\Phistl$
    is shifted by $\pi$ in order to overlay W$^+$ and W$^-$ 
    distributions in the same plot.
    2. The jet with $0\leq\Phistj\leq\pi$ is arbitrarily chosen as the quark
    (antiquark) jet from the decay of the \Wm\ (\Wp). 
}
 \label{fig:angdist} 
\end{figure}

In Figure~\ref{fig:angdist} (a-e) we show the distributions of all the five
angles obtained from the combined \Qqln\ event sample,
 and the expected distributions for $\dgz = \pm 1$ and 0.
These expected distributions are obtained from fully simulated
\MC\ event samples generated with \Excalibur, normalised
to the number of events observed in the data. Sensitivity to \tgc\
is observed mainly for \Cthw. The contribution of \Cthstl, \Phistl, 
\Cthstj\ and \Phistj\ to the overall
sensitivity enters mainly through their correlations with \Cthw. 


\subsection{The Optimal Observable (OO) analysis}   
\label{subs:oo}                          
The concept of optimal observables~\cite{OO} is used to
project the five kinematic variables of each event onto a
single observable. It has been shown that for a differential \xse\
which is linear in the parameter to be determined, the
sensitivity of the optimal observable is the same as for a 
multi-dimensional maximum likelihood fit. The W-pair differential \xse\ 
is, however, a second-order polynomial in the couplings, thus leading to
a loss of sensitivity when constructing the optimal observable without 
the quadratic term. Nevertheless,
in the case where the deviations of the couplings from zero are small,
the quadratic terms are suppressed and 
the loss in sensitivity is far outweighed by the 
great advantage of a parametrisation in only one dimension. This method 
also allows usage of all the available kinematic information from the 
event, whereas in other methods (see the next sub-sections)
some part of the information is lost due to computational limitations.

The optimal observable for measuring any particular TGC parameter, \al, 
is constructed for each event $i$ with the set of phase-space variables 
$\Omega=(\Cthw,\Cthstl,\Phistl,\Cthstj,\Phistj)$ by
differentiating the differential \xse\ for the event,
$\sigma_i(\al)\equiv d\sigma(\Omega,\al)/d\Omega)|_{(\Omega=\Omega_i)}$,
with respect to the parameter \al\ evaluated
at the Standard Model value, \al=0, and normalising to the
Standard Model \xse,
\[
 \OO_i =
\frac{1}{\sigma_i^{\mathrm{SM}}} \frac{d\sigma_i(\al)}{d\al}|_{(\al=0)}.
\]
The differential \xse\ used here is the Born \xse,
taking into account the twofold ambiguity in the definition of 
the measured \Cthstj\ and \Phistj.

Figure~\ref{fig:angdist}(f) shows the distribution
of the optimal observable corresponding to \dgz, along with the 
predictions corresponding to the \SM\ and to $\dgz=\pm 1$, as
obtained from corresponding Monte Carlo samples. Similar distributions 
are obtained for the optimal observables corresponding to \dkg\
and \lg.  

A binned maximum likelihood fit of the expected optimal observable
distributions to the data is performed to extract each coupling,
assuming the other two couplings to have \SM\ values. The expected optimal  
observable distributions for various couplings are derived from 
\MC\ events and normalised to the number of events in the data
in order to exclude any information from the overall production 
rate in this part of the analysis.  
To obtain the optimal observable distribution at intermediate parameter 
values, a reweighting technique is applied to sets
of Monte Carlo events generated with different couplings using 
a Born-level differential \xse\ but taking into account ISR effects.
 
The method is tested with about 350 Standard Model and 200
non-Standard Model Monte Carlo subsamples, each corresponding to the
collected luminosity. Summing up the \LL\ curves from the 
different subsamples corresponding to the same coupling gives 
results which are consistent with the coupling value used in 
the Monte Carlo generation. Unfortunately, for the \SM\ subsamples
the \LL\ curves corresponding to \dkg\ tend to have two minima, one 
close to the generated value, \dkg=0, and the second around \dkg=2.5. 
The second minimum is due to the 
fact that, for example, the normalised differential \xse\ for 
\dkg=2.5 is more similar to the \SM\ than the one for \dkg=1. 
In about 18\% of the cases the wrong minimum is the deepest one.
However, the \xse\ information already
excludes a value around 2.5 for \dkg\ (see Figure~\ref{fig:rateshape}). 
Therefore, the minimum at that point can be disregarded.
A similar problem occurs in 3\% (7\%) of the \SM\ subsamples
also for \dgz~(\lg) where the constraint from the \xse\
information is even more stringent, so it is handled in the same way.  

The error interval is defined in the usual way, as the region where
the \LL\ function is higher than its minimum by no more than 0.5.
To test the reliability of this error estimate, the fraction of 
subsamples where the correct value is within the error
interval is calculated. The error estimate is considered to be reliable
if the calculated fraction is consistent with 68\%, 
which is the case for all couplings.

Performing the fit to the 183~\GeV\ data, we obtain the results quoted in the 
first row of Table~\ref{tab:ooresult}. The statistical errors obtained
from the fit are consistent with the expected values which are 
estimated from the distribution of the Monte Carlo subsample fit results. 

Biases introduced by uncertainties in the simulation of the detector 
acceptance are
checked by removing events in phase-space regions where the
efficiency changes rapidly. Cuts on the polar angles of the charged
lepton and the neutrino, on the angle between the lepton and the nearest 
jet, as well as a cut on the lepton energy, were introduced. 
Furthermore the range of the OO-values that are considered in the fit
is changed by approximately $\pm 10\%$.
The variations of the fit result due to these
changes fall within the statistical fluctuations determined 
using many Monte Carlo subsamples with the size of the data sample.

The effect of \MC\ statistics is studied by allowing the expected
number of events in each bin of the optimal observable distribution 
to vary around its central value within its error according to 
a Gaussian distribution. This effect is found to be negligible.

Additional systematic uncertainties are studied due to the following 
effects.
\begin{itemize}
\item[a)] For quark jets in the \MC\ samples, the resolutions of the three 
  jet parameters (energy, $\cos\theta, \phi$)
  are varied by 10\%, and the energy scale is shifted by 0.5\%
  to account for systematic uncertainties in the modelling of
  the jet reconstruction. The
  size of the variations is determined from extensive studies of
  back-to-back jets at \LepI\ \Zz\ energies. A possible systematic shift in
  the reconstructed direction of the W boson has been estimated using
  radiative \ZGqq\ events. The shift in $|\cos\theta_{\rm W}|$ was
  found to be less than 0.01~\cite{tgc172-analysis}.
  The same reconstruction uncertainties are assumed also for the $\tau$
  jets in \Qqtn\ events.
\item[b)]Uncertainties in the fragmentation model are studied using
  a Monte Carlo sample generated with \Herwig\ rather than \Pythia.
\item[c)]Differences between Monte Carlo generators not related to
  fragmentation are investigated by replacing the \Excalibur\ 
  reference sample by \Koralw. This latter generator also uses the \Jetset\ 
  fragmentation scheme but has a different treatment of ISR and simulates
  correctly the $\tau$ helicity effects.
\item[d)]Systematic effects arising from the simulation of the \ZGqq\
  background are taken into account by replacing \Pythia\  
  with \Herwig. The two-photon background is removed to test
  its impact on the measurement. The four-fermion processes from
  \Wenu, \Zee\ and \ZZ\ are added as background to the CC03 diagrams,
  neglecting interference. The justification for this
  simplification is verified by using a fully simulated four-fermion
  Monte Carlo (\Excalibur) as the reference sample.
  In addition, subsamples of fully simulated
  four-fermion Monte Carlo events have been fitted and no bias was
  found. To test the Monte Carlo generators, samples of \Koralw\ and
  \Grace\ generated with the CC03 W-pair diagrams according to the
  \SM\ are used as test samples and all the TGC results are found to
  be consistent with zero.
\end{itemize}

\begin{table}[htbp]
\begin{center}
\begin{tabular}{|ll|r|r|r|}\hline 
  \multicolumn{2}{|c|}{Source}&  \multicolumn{3}{c|}{Error on parameter}  \\
  \multicolumn{2}{|c|}{ } & \dkg   &  \dgz   &  \lg  \\ \hline 
  a) & Jet reconstruction     & .114   & .029    & .033 \\
  b) & Fragmentation          &\dabfhe &\ddgzhe  &\dawhe \\
  c) & MC generator           &\dabfmc &\ddgzmc  &\dawmc \\
  d) & Background             & .093   & .029    & .010  \\ \hline
     & Combined               &\dabfsu &\ddgzsu  &\dawsu \\ \hline 
\end{tabular}
\end{center}
\caption{ Contributions to the systematic uncertainties in the 
  determination of the \tgc\ parameters in the OO analysis for the 
  \WWqqln\ angular distributions of the 183~\GeV\ data. }
\label{tab:oosys}
\end{table}

Table~\ref{tab:oosys} summarises the various components of the systematic 
errors. These are incorporated into the \LL\ functions by 
convolving their distributions, assumed to be Gaussian, with the 
likelihood functions. The corrected likelihood functions resulting
from this convolution are
used to obtain new values for the TGC parameters with modified 
errors which include the systematic uncertainties. The new TGC values 
may differ from those obtained before the convolution due to the 
asymmetric nature of the likelihood function. 
The results are listed in Table \ref{tab:ooresult}. 


\subsection{The Binned Maximum Likelihood (BML) analysis}   
\label{subs:bml}
The BML analysis has been used to analyse our previous data taken
at 161~\GeV\ and 172~\GeV\ and is fully described in 
\cite{tgc161-analysis,tgc172-analysis}. The finite statistics of the data 
and \MC\ samples limits the number of kinematic observables which can be 
used to three. The most sensitive ones, namely \Cthw, \Cthstl\ and
\Phistl\ are chosen\footnote{
The observables \Cthstj\ and \Phistj\ have smaller sensitivities
due to their twofold ambiguity.}. The expected
three-dimensional distribution of these observables as a function of 
the TGCs is fitted to the corresponding distribution of the data.
The expected distribution is obtained from MC events
generated according to the \SM\ with the CC03
diagrams. These events are reweighted 
to correspond to any other TGC value. The reweighting procedure uses 
very large statistics MC samples at the generator level, generated with      
ten different sets of TGC values and used to produce ten basis 
distributions. Since the differential \xse\ is a second-order
polynomial in the \tgc s (see section~\ref{subs:tgcrate} above), 
the distribution corresponding to any other set 
of TGC values can be obtained by a proper linear combination
of the ten basis distributions. 

This procedure accounts for the effects of efficiency, resolution and 
background from other \WW\ decay channels. To account also for
non-\WW\ background, the contribution of background sources which do 
not depend on the TGCs, namely \ZGqq, \Qqee\ and two-photon reactions, 
is added to the expected distribution used in the fit.
The distributions from these sources are obtained from corresponding
\MC\ samples. Background from other four-fermion diagrams 
which is TGC-dependent is neglected at this stage, and biases  
caused by this neglect are evaluated by performing the 
fit on four-fermion MC events taken as a test sample. These biases are 
subtracted from our fit results.

The expected distribution is normalised to the number of 
events observed in the data in order not to incorporate any
information from the overall production rate in this part of the analysis.
The probability for observing the number of events seen in each bin  
is calculated using Poisson statistics. 
The statistical fluctuations in the Monte Carlo
are taken into account using the method of reference~\cite{BARLOW}.
Consequently, the errors obtained from the fit include both data and
\MC\ statistics. 

Detector level MC samples generated with different TGCs are used
to verify that the BML fit method introduces no bias.
The reliability of the statistical errors obtained from the fit is
checked with many \MC\ subsamples, in the way described in 
section~\ref{subs:oo}. A problem occurs only for \dkg\ where 
the fraction of subsamples with the correct TGC value within the
error interval falls below 68\%. It is found that dividing the \LL\ 
function by a correction factor of 1.21 increases the error interval
in such a way that this fraction
reaches 68\%. This correction factor is then applied to the
\LL\ function extracted from the data.
The statistical errors obtained in this way are also consistent 
with the expected values which are listed in Table~\ref{tab:ooresult}.

The systematic errors are evaluated and incorporated into the
\LL\ functions in a similar way as for the OO method.
The results are consistent with those of the OO method. 


\subsection{The Spin Density Matrix (SDM) analysis}   
\label{subs:sdm}                          
Spin density matrix elements are observables 
directly related to the polarisation of the W bosons~\cite{GOUNARIS,LEP2YR}.
Additional insight into the underlying physics may be gained from 
these spin-related observables, and the relative production 
of various helicity states of the W bosons can be measured.
Comparing the spin density matrix elements with the theoretical 
predictions allows a model-independent test of the TGCs.
If deviations were detected, this method would give information on the 
structure of the anomalous couplings.

Spin density matrix elements are normalised products
of the helicity amplitudes $\mathcal{F}_{\tau_-\tau_+}^{(\lambda)}(\x)$,
where $\tau_-$ and  $\tau_+$ are the helicity states of the $\Wm$ and 
$\Wp$ boson, respectively, and $\lambda$ denotes the spin of the 
$\epem$ system.
The two-particle joint density matrix elements are defined according to
\[
 \rho_{\tau_-\tau'_- \tau_+\tau'_+ }(\x) = 
   \frac{  \sum_{\lambda}
     \mathcal{F}_{\tau_-\tau_+}^{(\lambda)}( \mathcal{F}_{\tau'_-\tau'_+}^{(\lambda)})^{\ast}}
{\sum_{\lambda \tau_+ \tau_-} | {\mathcal F}_{\tau_-\tau_+}^{(\lambda)}|^{2}}.
\]
Due to limited statistics in the analysis described here the single-W 
spin density matrix elements $\rho_{\tau \tau'}$ are used. 
The spin density matrix of the \Wm\ boson, for example, is obtained from 
the two-particle joint density matrix elements by the relation 
\[
 \rho_{\tau_-\tau_-'}^{\scriptsize \Wm}(\x) =  
      \sum_{\tau_+} \rho_{\tau_-\tau'_- \tau_+\tau_+ }(\x).
\]
The matrix $\rho_{\tau \tau'}$ is hermitian, thus having six independent 
matrix elements. 
The diagonal elements $\rho_{\tau \tau}$ of the spin density matrix are
real and can be interpreted as the 
probability to produce a W boson with helicity $\tau$.
The off-diagonal elements are complex in general, but 
for CP-conserving theories the imaginary parts vanish.

The \sdmes\ are extracted with the help of projection 
operators~\cite{GOUNARIS,LEP2YR}. These operators reflect the standard 
$V-A$ couplings of the fermions to the W boson in the W decay.
The single-W  \sdmes~ $\rho_{\tau \tau'} $
 can be extracted using the threefold
differential \xse\ $d\sigma$/\dreidiff\ from the relation
\beq
\label{eq-poptheo}
\rho_{\tau \tau'}^{\scriptsize {\rm W_1}}(\x) 
\frac{d \sigma(\epem \rightarrow {\rm W_1 W_2})}{d \x} = 
\frac {1}{{\rm Br_{\rm W_1 \rightarrow f\bar{f}}}}
\int \frac{d \sigma(\epem \rightarrow {\rm W_2 f\bar{f}})}
{d\x d\cos\theta^*_1 d\phi^*_1}
\Lambda_{\tau \tau'}(\theta^*_1,\phi^*_1) d\cos\theta^*_1 d\phi^*_1.
\eeq
Here $\Lambda_{\tau \tau'}$ is the suitable projection operator for
extracting the \sdme~ $\rho_{\tau \tau'}$ and 
${\rm Br_{\rm W_1 \rightarrow f\bar{f}}}$ is the branching ratio
for the decay of the W boson considered.
Expressions for the projectors can be found in~\cite{GOUNARIS}.
Experimentally, equation \ref{eq-poptheo} corresponds to 
\[
\rho_{\tau \tau'}(\x)= \frac{1}{N} 
      \sum_{i=1}^{N} \Lambda_{\tau \tau'}(\y_i,\z_i)
\]
where N is the number of events and 
$\Lambda = \Lambda(\y_i,\z_i)$ is the value of the projection operator
of event $i$.
The spin density matrix elements are extracted in bins of \x.

For the leptonically decaying W, the full information about the 
W decay angles is accessible and the matrix elements can be directly
extracted according to (\ref{eq-poptheo}).
For hadronic decays of the W boson where the quark cannot be 
distinguished from the antiquark, only the folded angular distributions of
\y\ and \z\ are directly available. These
folded distributions can be identified only with the symmetric\footnote{
Symmetric/antisymmetric under the transformation 
\y\ $\rightarrow$ --\y, \z\ $\rightarrow$ \z$+\pi$}
part of the angular distributions~\cite{BILENKY}, whereas no information 
about the antisymmetric part is accessible.
The following symmetric (combinations of) spin density matrix elements 
can be extracted from the folded angular distribution
by using the symmetric part of the projection operators only:
$$ \rho_{++} + \rho_{--},~~ \rho_{00},~~ {\rm Re}(\rho_{+-}),
~~ {\rm Im}(\rho_{+-}),~~ {\rm Re}(\rho_{+0} - \rho_{-0}),~~
{\rm Im}(\rho_{+0} + \rho_{-0}).$$ 
The correlations between the single-W density matrix elements are calculated
analytically. 
 
In order to compare the \dmes\ extracted from data with the theoretical
predictions, the data have to be corrected for experimental effects, like
 selection efficiency, acceptance, angular resolution and the
effect of ISR. A simple approach,
based on a Standard Model Monte Carlo, is performed.
First the expected background, taken from Monte Carlo, is subtracted on 
a statistical basis. Subsequently the data are corrected by multiplying the
three-dimensional angular distribution by
a correction function which is given by the ratio of the 
angular distributions for the reconstructed angles in the 
selected events and the generator level distributions of all events.
The effect of ISR is approximately accounted for by reducing the 
centre-of-mass energy by the mean energy of initial state photons, 
$\langle E_{\rm  ISR} \rangle = 1.6$~\GeV, as determined from Monte Carlo.

\begin{figure}[tbhp]
\epsfig{file=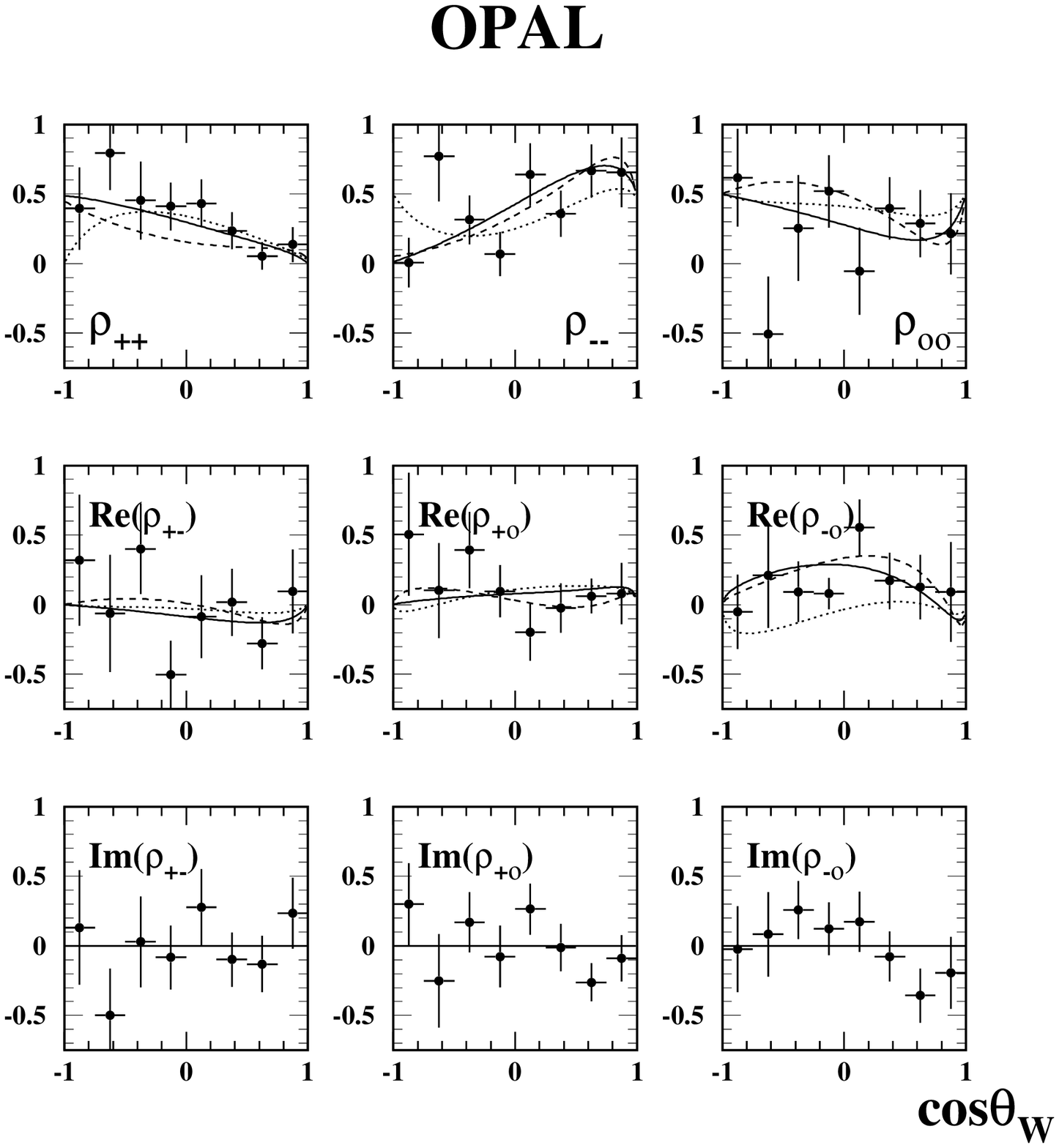,width=\textwidth}
   \caption{\sl
    Spin density matrix elements for the leptonically decaying W
    as a function of \Cthw. The data points are corrected for 
    experimental effects. The solid (dotted, dashed) lines
    show the predictions of models with \dgz=0 (+1, --1). All
    other anomalous couplings are set to zero.
}
 \label{fig:sdmlep} 
\end{figure}

\begin{figure}[tbhp]
\epsfig{file=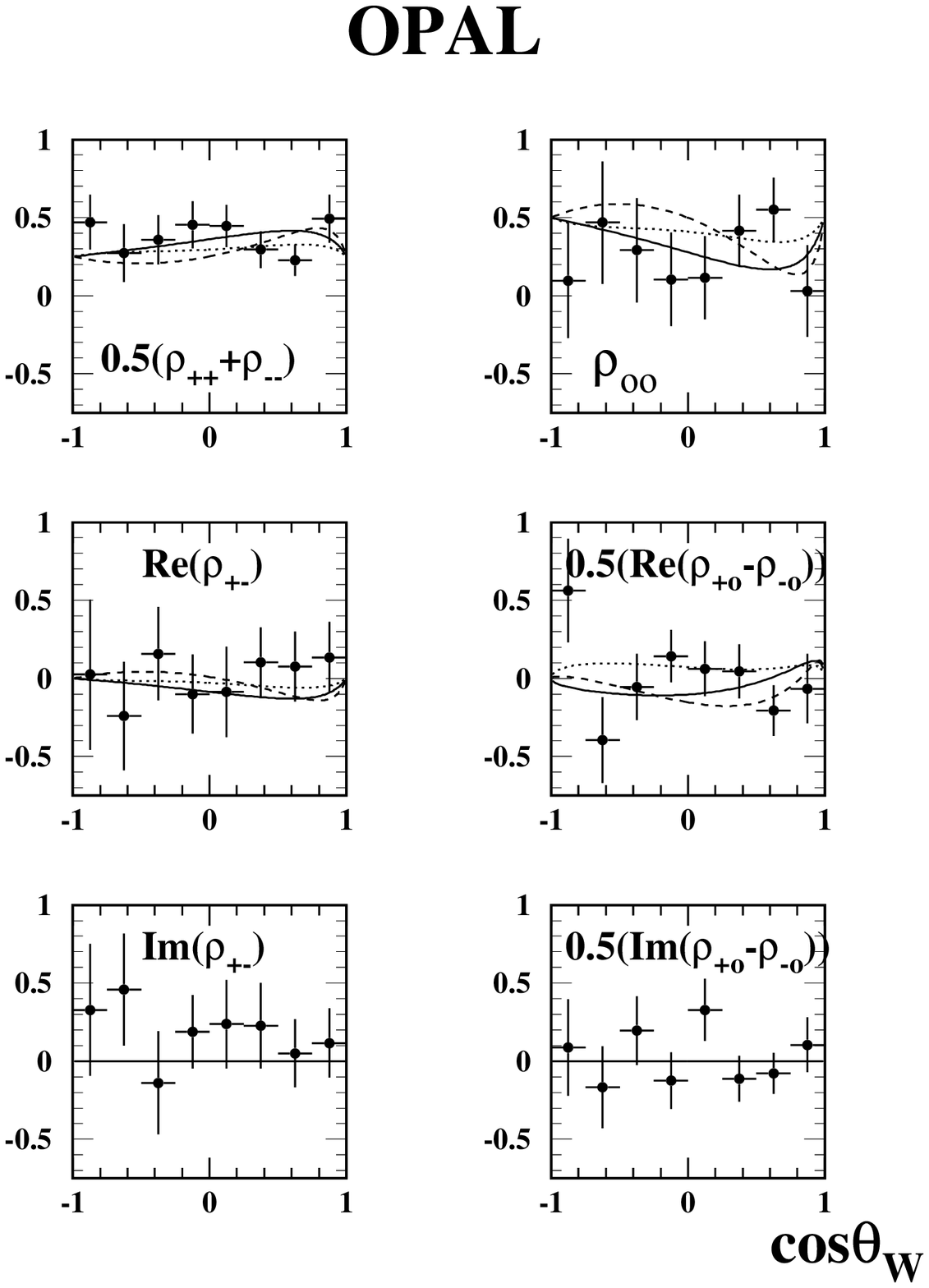,width=\textwidth}
   \caption{\sl
    Spin density matrix elements for the hadronically decaying W
    as a function of \Cthw. The data points are corrected for  
    experimental effects. The solid (dotted, dashed) lines
    show the predictions of models with \dgz=0 (+1, --1). All
    other anomalous couplings are set to zero.
}
 \label{fig:sdmhad} 
\end{figure}

Distributions of
the spin density matrix elements of the leptonically decaying W and the 
hadronically decaying W 
are shown in Figures~\ref{fig:sdmlep} and \ref{fig:sdmhad}.
Overlaid are the analytical predictions as expected in the Standard 
Model. The errors are statistical. The systematic errors
are estimated to be less than 10\% of the statistical errors. 
This is verified by replacing the
\SM\ Monte Carlo sample used for the correction of detector effects
with samples generated with $\dgz=\pm 1$
values which are far outside the allowed region obtained in our fit 
(see below).

\begin{figure}[tbhp]
\epsfig{file=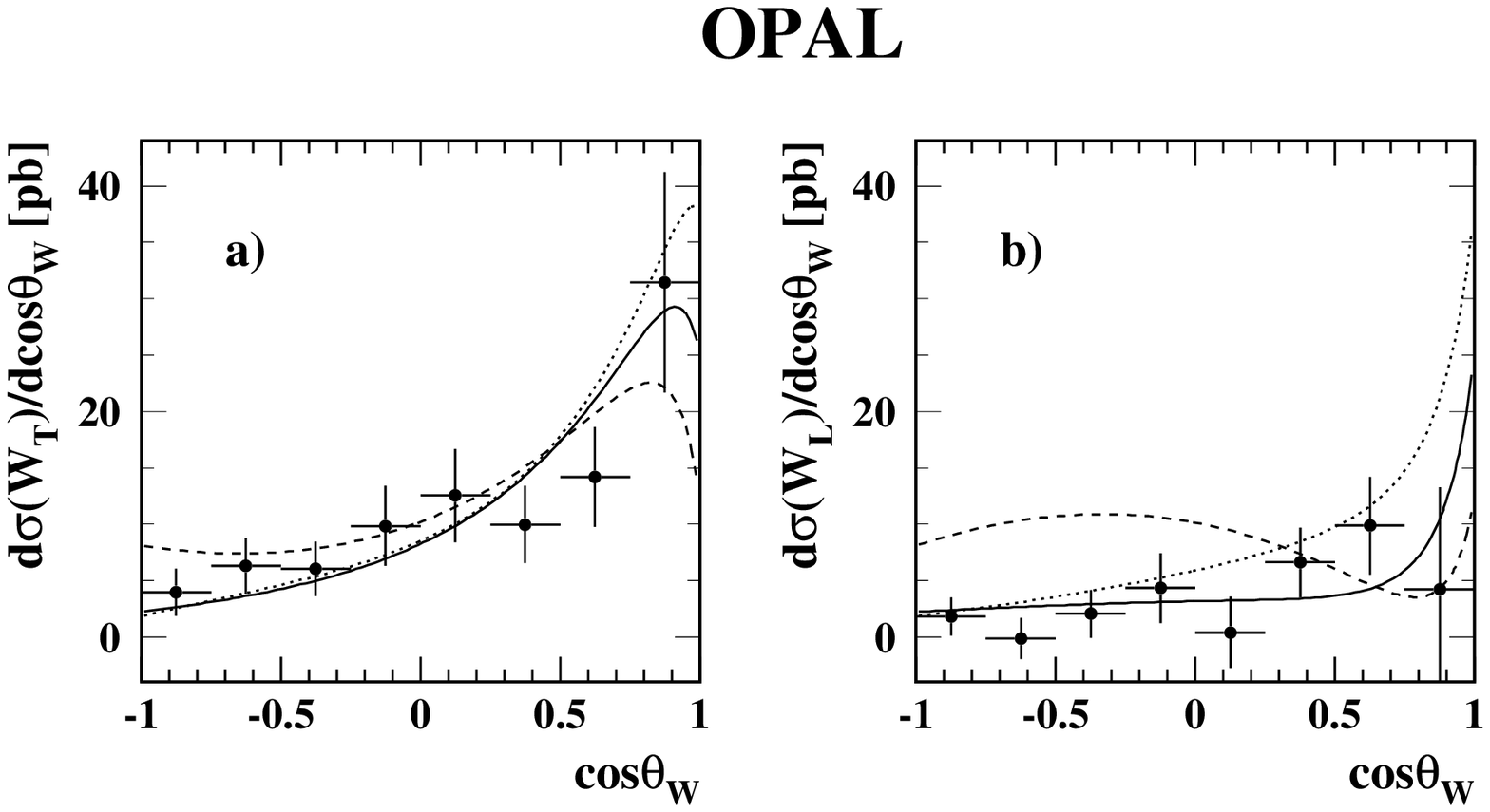,bbllx=5pt,bblly=250pt,bburx=520pt,bbury=530pt,
width=\textwidth,clip=}
   \caption{\sl
    Differential \xse\ to produce a) a transversely polarised 
    W and b) a longitudinally polarised W in a W-pair event where the
    second W can have any polarisation. The points represent the 
    data and the solid (dotted, dashed) lines show the predictions of
    models with \dgz=0 (+1, --1). The error
    bars include statistical and systematic uncertainties, except
    for a normalisation error of 4.3\% associated with the 
    total \xse\ measurement.  
}
 \label{fig:dsdcos_lt} 
\end{figure}

The measurements of the W-pair \xse, the \Cthw\ distribution and 
the spin density matrices can be combined to give a measurement of the 
semi-inclusive differential \xses, to produce a transversely
polarised W, \epem\ra\WT W, or a longitudinally polarised W, \epem\ra\WL W,
where in either case the second W can have arbitrary helicity. For this 
purpose, the raw \Cthw\ distribution, plotted in 
Figure~\ref{fig:angdist}(a), is corrected 
for detector effects by subtracting the background and multiplying by
a correction function for efficiency and resolution obtained from \SM\ 
\MC. The systematic error associated with 
this correction is estimated in the same way as for the OO analysis.
In addition, as was done for the spin density matrix elements, the \SM\
Monte Carlo sample used for the correction of detector effects is replaced
by samples generated with $\dgz=\pm 1$.
The corrected \Cthw\ distribution is multiplied by
our measured total \xse\ and the corresponding spin density
matrix elements after combining their values from the leptonic and
hadronic decays. The resulting differential \xses, plotted
in Figure~\ref{fig:dsdcos_lt}, are seen to be consistent with the Standard 
Model expectations. Integrating over these \xses, the overall
fraction of longitudinally polarised W bosons is determined to be 
0.242$\pm$0.091$\pm$0.023, where the systematic error is dominated by
uncertainties in the jet and tau resolutions (0.017) and MC generator 
(0.015). The expected value for this fraction is 0.272 (0.392, 0.405)
for the \SM\ (\dgz=+1, --1). 

The spin density matrix elements can be used to 
extract the TGCs by comparing them, before the various 
corrections, to those expected from Monte Carlo events for different
TGC values. This is done with a reweighting technique, taking into
account all experimental effects. The W production angle and the
spin density matrix elements of both W bosons are used in the fit. 
We obtain results consistent with those from the OO method.


\subsection{Summary of the \Qqln\ analyses}   
\label{subs:qqlnsum}
                          
All three methods described above give consistent results 
for the 183~\GeV\ data. 
The OO results are chosen to be
combined with the results of the other channels and the \xse\
result since this method is expected to give the best sensitivity, 
as can be seen
from the expected errors listed in Table~\ref{tab:ooresult}.

\begin{table}[bht]
\begin{center}
\begin{tabular}{|l|c|c|c|} \hline 
Level of results      & \dkg\    & \dgz\  &  \lg\   \\ \hline 
& & & \\
Without systematics   & \abfsta & \dgzsta & \awsta \\
& & & \\
Expected errors, OO method  & \abfexp & \dgzexp & \awexp \\
Expected errors, BML method  & $\pm 0.56$ & $\pm 0.16$ & $\pm 0.19$ \\
Expected errors, SDM method  & $\pm 0.50$ & $\pm 0.14$ & $\pm 0.16$ \\
& & & \\
Including systematics & \abfold & \dgzold & \awold \\ 
& & & \\
Including 161/172~\GeV\ angular data & \abftyr & \dgztyr & \awtyr \\
& & & \\
Including \xse\ & \abfbot & \dgzbot & \awbot \\
& & & \\ \hline 
\end{tabular}
\caption{Measured values of the TGC parameters using the \Qqln\
event selection analysed with the OO method, with and without taking 
into account the systematic uncertainty. We also list the results
after combining with our 161 and 172~\GeV\ data and after combining  
with the \xse\ information of \Qqln\ data from all \Com\ 
energies. The expected statistical errors obtained from Monte Carlo
studies are listed for all three analysis methods.}
\label{tab:ooresult}
\end{center}
\end{table}

Although our method of OO analysis using the optimal observable
distribution is preferred for single parameter fits, it is less well
suited to a fit of two or three \tgc\ parameters. In this case,
two- or three-dimensional distributions of the optimal observables
corresponding to the different \tgc\ combinations would be required,
and it is simpler to fit the angular variables directly, using the 
BML method.  

The 183~\GeV\ results are combined with our 
161~\GeV\ and 172~\GeV\ results~\cite{tgc172-analysis}. In the
161~\GeV\ analysis~\cite{tgc161-analysis} only the parameter \awf\
has been analysed. Therefore, this analysis is extended here using the
same (BML) method to include the three parameters \dkg, \dgz\ and \lg,
along with the two-dimensional and three-dimensional fits. All the tests
and the systematic studies are done in the same way as for the 183~\GeV\
BML analysis. The correlation between the systematic errors of the 
\Qqln\ results from the three \Com\ energies is neglected since,
using the BML method, it is found to affect the results by no more than 
10\% of their statistical errors.

\begin{figure}[htb]
\epsfig{file=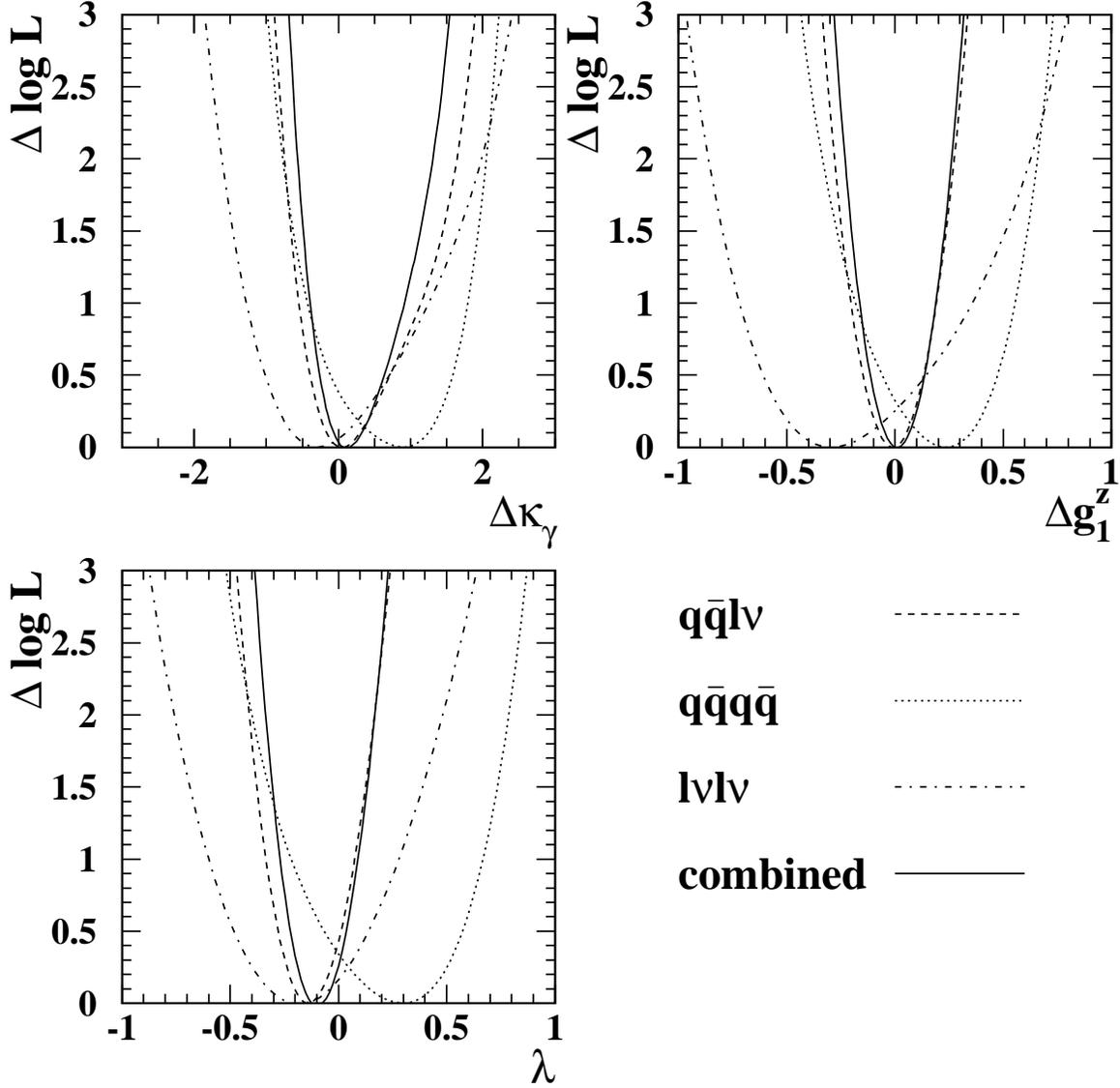,width=\textwidth}
\caption{\sl
Negative log-likelihood curves obtained in the different event  
selections: \Qqln\ (dashed lines), \Qqqq\ (dotted lines) and \Lnln\
(dash-dotted lines). Each curve is obtained by combining the results
from the angular distributions and the \xse\ and setting
the other two \tgc\ parameters to zero. Systematic
errors are included. The solid line is obtained by combining the 
three event selections.
}
\label{fig:channel}
\end{figure}

Table~\ref{tab:ooresult} summarises the \Qqln\ results from the 
183~\GeV\ data, 
the results obtained after combining the 161 and 172~\GeV\ data,
and the combined \Qqln\ results obtained after adding also the 
\xse\ information for that channel from all \Com\ energies. 
The \LL\ curves corresponding to the combined \Qqln\ results are shown
in Figure~\ref{fig:channel}. The correlation between the systematic errors
of the angular distributions and the \xse, mainly due to the
uncertainty in the background level, is neglected as it affects the 
results by less than 1\% of their statistical errors.


\section{Analysis of the \WWqqqq\ angular distributions}
\label{sec:qqqq}
The analysis of the \WWqqqq\ channel is performed on both the 172~\GeV\ 
and the 183~\GeV\ data. It uses the same 
event samples as the \xse\ analyses (see 
section~\ref{subs:qqqqsel} and ref.~\cite{tgc172-analysis}).

\subsection{Event selection and reconstruction}
\label{subs:qqqqrec}
Using the Durham $k_\perp$ algorithm \cite{DURHAM}, each selected event 
is forced into 4 jets, whose energies are   
corrected for the double counting of charged track momenta and 
calorimeter energies~\cite{GCE}.
To improve the resolution on the jet four-momentum we perform  
a kinematic fit requiring energy-momentum conservation and
equality of the masses of the two W candidates (5-C fit).
The event reconstruction is complicated 
by the ambiguity  in the choice of the correct di-jet combination 
and uncertainties in the determination of the W charge. 
The latter is assigned by comparing
the sum of the charges of the two jets coming from the same W candidate.
Each W charge is defined as:
\[
Q_{W_1}=\frac{N_i+N_j}{\sum_{m=1}^4D_m},
\,\,\,\,\,\,\,\,\,\,\,\,\,\,\,\,\,\,\,
Q_{W_2}=\frac{N_k+N_l}{\sum_{m=1}^4D_m},
\]
where jets $i$ and $j$ belong to W$_1$,  jets $k$ and $l$
belong to W$_2$, and $N_{jet}$ and $D_{jet}$ are given by:
\[
{N_{jet}}={\sum_{i=1}^{N}q_i|p_{i||}|^{0.5}},
\,\,\,\,\,\,\,\,\,\,\,\,\,\,\,\,\,\,\,
{D_{jet}}={\sum_{i=1}^{N}|p_{i||}|^{0.5}}
\]
where $q_i$ is the charge of the $i^{\rm th}$ track, $p_{i||}$ is the 
projection of its momentum along the jet axis and $N$ 
is the total number of tracks in the jet.
 
\begin{figure}[hbt]
\epsfig{file=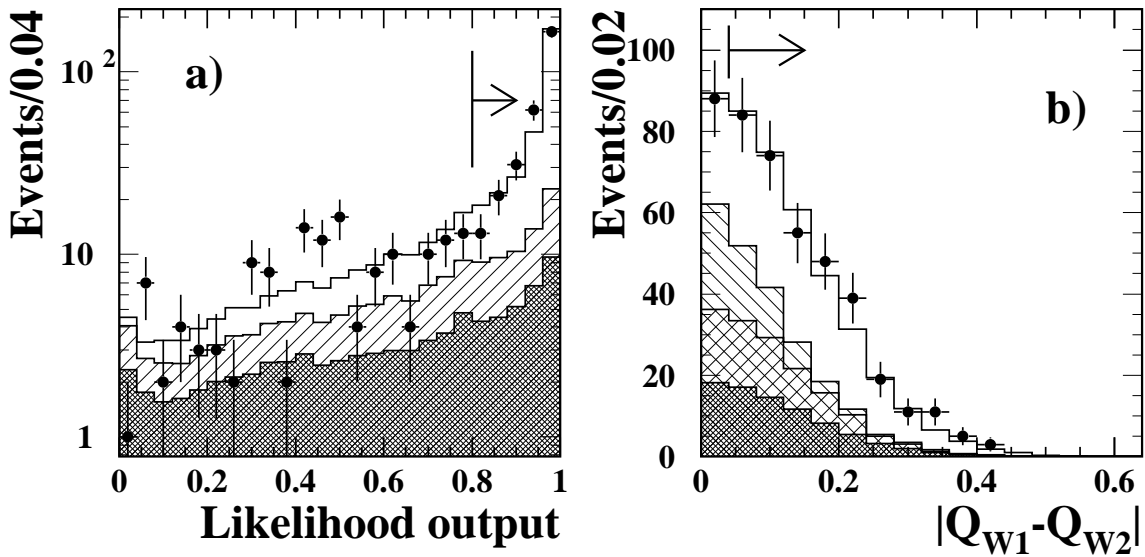,width=\textwidth}
\caption{\sl
 For the \Qqqq\ channel,
 a) distribution of the output of the jet-pairing likelihood corresponding 
 to the most likely combination for the data collected at 183~\GeV\ (points) 
 and for the Monte Carlo (histogram). The hatched area shows the 
 contribution of wrong pairing, and the dark area represents the 
 contribution of the background. The arrow indicates the cut value.
 b) Distribution of the charge separation of the W candidates for the data 
 collected
 at 183~\GeV\ (points) and for the Monte Carlo (histogram). The hatched 
 area shows the contribution of correct pairing and incorrect W charge,
 the double hatched area shows the contribution of wrong pairing, and 
 the dark area represents the contribution of the background.
 The arrow indicates the cut value.
}
\label{fig:cutlikQw}
\end{figure}

\begin{figure}[hbt]
\epsfig{file=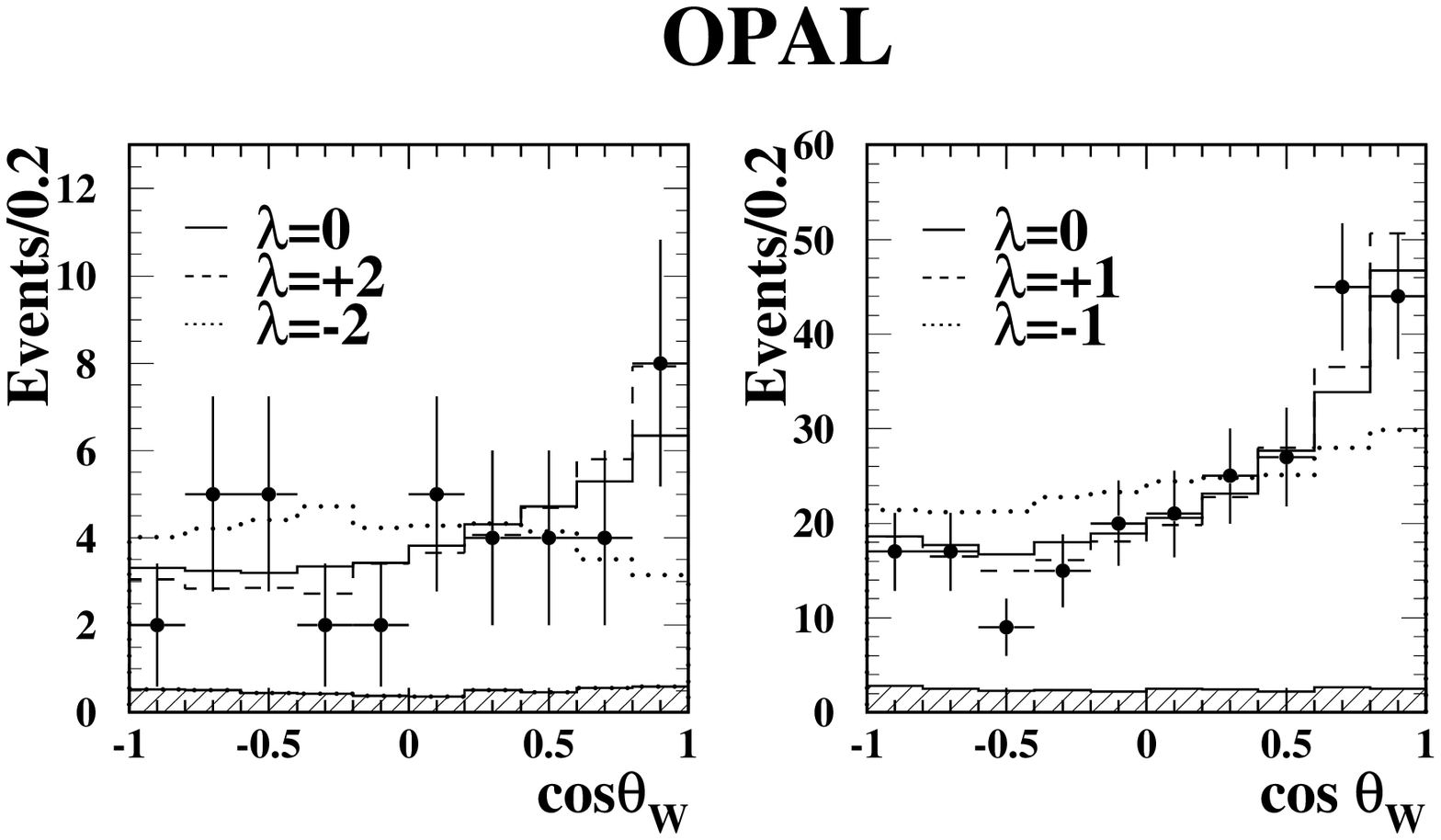,width=\textwidth}
\caption{{\sl
Distribution of \Cthw\ in \Qqqq\ events as obtained at 172~\GeV\ (left) 
and 183~\GeV\ (right). The solid points
are the data. The Monte Carlo predictions for \lg=0,+2,--2 and
\lg=0,+1,--1 are shown at 172 and 183~\GeV, respectively, as solid, dashed 
and dotted lines. The hatched histograms show the contributions
of the \ZGqq\ background, as predicted by} \Pythia.}
\label{fig:costW}
\end{figure}

One di-jet combination is chosen out of the three possible
ones by a likelihood algorithm. The input variables to the likelihood
are the di-jet invariant masses, obtained by a kinematic
fit requiring energy and momentum conservation (4-C fit), 
the charges of the two W candidates, and the 
probabilities of the 5-C kinematic fits.
Once a jet combination is chosen, the \Wm\ is defined
as the di-jet whose charge is more negative.
According to the Monte Carlo simulation,
these choices correspond to a probability of selecting the correct di-jet 
combination of about 78$\%$ (82$\%$) at 183 (172)~\GeV, and to 
a probability of correct assignment of the W charge, once the correct
pairing has been chosen, of about 76$\%$. 
Both probabilities show a few percent dependence on the 
value of an eventual anomalous \tgc.
Incorrect jet pairing and wrong determination of the W charge decrease
the sensitivity of the reconstructed \Cthw\ distribution
to possible \tgc s. 
To increase the fraction of events correctly
reconstructed, we apply further cuts on the value of the 
jet pairing likelihood, $L > 0.8$, and on the W charge separation,
$|Q_{{\rm W}_1}-Q_{{\rm W}_2}|> 0.04$. The distributions of these variables,
for the data collected at 183~\GeV\ and for the 
corresponding Monte Carlo, are shown in
Figure~\ref{fig:cutlikQw}. The cut values are the result of
a compromise between loss in efficiency and gain in sensitivity.
After these cuts, the probability of correct
pairing increases to 86\% (87\%) at 183 (172)~\GeV\ and the
probability of correct determination of the W charge to 82\%. 
The selection efficiency is reduced  by about 40\%, 
whereas  the \ZGqq\ 
background is reduced by 60\% and 70\% at 172 and 183~\GeV\, respectively. 
At 172~\GeV, 41 events survive all cuts out of the 59 initially
selected, and 241 out of 438 events remain finally at 183~\GeV.
The \Cthw\ distributions obtained at both energies are shown in Figure
\ref{fig:costW} together with the distributions predicted by 
the Monte Carlo for different values of \lg.


\subsection{TGC analysis}
\label{subs:qqqqtgc}

Due to lack of separation between
quarks and antiquarks, in \WWqqqq\ events most of the
sensitivity  to the TGCs is contained in the \Cthw\ distribution.
A binned maximum likelihood method is thus used in this channel. 
The measured \Cthw\ distribution is divided into ten bins in the 
range [--1,1]. The expected \xse\ in each bin $i$, $\sigma_i$,
due to \WW\ production is parametrised  as a second-order polynomial 
in the analysed coupling. Monte Carlo reference histograms, 
including all effects of acceptance, resolution, incorrect jet pairing
and incorrect determination of the W charge, are used to build the 
parametrisation. The contribution of the \ZGqq\ background as obtained from
the \Pythia\ Monte Carlo is added to the W-pair expectation.
To avoid any dependence on the overall production rate, the expected
\xse\ is normalised to the total number of candidates. 
The expected distribution is fitted to the data using a binned
maximum likelihood method and assuming Poisson statistics.
The result of the fits to  
\dkg, \dgz\ and \lg\ as obtained from the 183~\GeV\ data sample 
are given in the first row of Table~\ref{tab:qqqqres} where the
quoted errors are statistical only.

The reliability of the statistical error estimates is checked by 
performing the analysis on many subsamples of Monte Carlo 
events, each corresponding to the
data luminosity. The r.m.s. values of the fit results corresponding to these
subsamples are quoted in the second row of Table~\ref{tab:qqqqres}.
These values are somewhat lower than the statistical errors coming
from the fits to the data. However, the data statistical errors are still
compatible with their expected values with probabilities above 5\%, as
obtained from the MC subsample study. As a further test, the 
fraction of subsamples where the correct TGC value turns out to be
inside the error interval is checked and found to be consistent with 
68\%, as required.

The following sources of systematic errors are considered:
\begin{itemize}
\item[a)]
The effect of possible differences in the jet resolution between data
and Monte Carlo is studied in the Monte Carlo by smearing and shifting 
the jet energies and directions in the same way as done for the \Qqln\ 
analysis. The resulting changes caused to the measured
\tgc s are added in quadrature and taken as a systematic error.
\item[b)]
Possible dependences on fragmentation models
are studied by comparing the results of the fit to two Monte Carlo
samples generated with \Pythia. In the first sample,
the \Jetset~\cite{PYTHIA} fragmentation model is implemented as
for the reference Monte Carlo samples, whereas
in the second it is replaced by \Herwig.
The implementation of \Herwig\ mainly results in a higher probability
of correct jet pairing and correct W charge with respect to \Jetset. 
The statistics of the data are insufficient to discriminate
between the two models. We assign as a systematic error for each
coupling the difference we find in the results of the fits to 
the two Monte Carlo samples.
\item[c)]
We study  possible biases due to the choice
of the Monte Carlo generator by repeating the fit to \WW\ samples
generated with  \Pythia, \Koralw\ and \Grace.
\item[d)]
A systematic error on the estimation of 
the background is determined by varying both its shape and 
normalisation. The shape predicted by \Pythia\ is replaced by that
predicted by \Herwig. The normalisation is 
varied by the background uncertainty as determined from the analysis
of the \xse.  The feed-through from other W boson decay channels
is less than 0.2\% and is neglected.
The effect of neglecting the contribution of four-fermion
diagrams is studied by performing the analysis on various 
four-fermion Monte Carlo samples generated with \Excalibur\ and \Grace.
The corresponding shifts on the fitted values 
of each TGC parameter are added in quadrature and taken as a
systematic error.
\item[e)]
Bose-Einstein correlations (BEC)
between bosons originating from different W in the
event might affect the measured W charge distribution. This effect
is studied at 183~\GeV\ with a \Pythia\ Monte Carlo sample~\cite{BEC} 
where BEC have been implemented.
\item[f)]
The measured W charge distribution might be affected also by colour 
reconnection. This effect is investigated at 183~\GeV\ with
Monte Carlo samples generated with \Pythia, in two classes
of reconnection models called type I and type II~\cite{RECONN},
and with \Ariadne~\cite{ARIADNE}. 
\item[g)]
Finally, a possible presence of biases in our analysis or fit method is 
checked by analysing 
Monte Carlo events generated with known values of the couplings, ranging
from --2 to +2. As a consistency check we use large Monte Carlo samples.
We also perform fits to many small Monte Carlo subsamples as mentioned 
above and examine the sum of their likelihood functions. 
\end{itemize}
 
\begin{table}[htbp]
\begin{center}
\begin{tabular}{|ll|c|c|c|}\hline 
\multicolumn{2}{|c|}{Source}&  \multicolumn{3}{c|}{Error on parameter}  \\
\multicolumn{2}{|c|}{ } & \dkg & \dgz & \lg \\ \hline 
a) &Jet reconstruction    & 0.26  & 0.08  & 0.09 \\
b) &Fragmentation         & 0.30  & 0.10  & 0.20 \\
c) &MC generator          & 0.45  & 0.10  & 0.13 \\
d) &Background            & 0.22  & 0.06  & 0.10 \\
e) &BEC                   & 0.07  & 0.04  & 0.02 \\
f) &Colour reconnection   & 0.01  & 0.01  & 0.01 \\
g) &Fit bias tests        & 0.20  & 0.08  & 0.09 \\
\hline 
   &Combined              & 0.68  & 0.19  & 0.29 \\ \hline 
\end{tabular}
\end{center}
\caption{
Contributions to the systematic errors
in the determination of the three \tgc $\:$ parameters 
for the \WWqqqq\ angular distribution analysis of the 183~\GeV\ data.}
\label{tab:qqqqsys}
\end{table}
  
The results of all these systematic studies for the 183~\GeV\ data are 
summarised in Table~\ref{tab:qqqqsys}. The systematic errors for the 
172~\GeV\ data are similar and are assumed to be fully 
correlated with those of the 183~\GeV\ data.
The combined systematic errors are convolved with the likelihood
functions. The results for the 183~\GeV\ data sample 
after convolving the systematic errors are
listed in the third row of Table~\ref{tab:qqqqres}. After combining
with the 172~\GeV\ data, we obtain the results listed in the fourth
row of Table~\ref{tab:qqqqres}. Finally, we combine also the results of
the \xse\ analysis in this channel, using the 161, 172 and 
183~\GeV\ data samples, yielding the results quoted in the last row
of Table~\ref{tab:qqqqres}. The corresponding \LL\ functions are plotted 
in Figure~\ref{fig:channel}. 

\begin{table}[htb]
\begin{center}
\begin{tabular}{ |l|c|c|c| } \hline 
Level of results      & \dkg\    & \dgz\  &  \lg\   \\ \hline 
& & & \\
Without systematics   & $1.15_{-1.25}^{+1.28}$ &
 $0.68_{-0.64}^{+0.91}$ & $0.76_{-0.67}^{+0.78}$ \\
& & & \\
Expected errors       & $\pm 1.13$ & $\pm 0.43$ &  $\pm 0.52$ \\
& & & \\
Including systematics & $1.16_{-1.34}^{+1.39}$ &
 $0.73_{-0.72}^{+0.88}$ & $0.79_{-0.78}^{+0.79}$ \\
& & & \\
Including 172~\GeV\ data & $1.15_{-1.36}^{+1.35}$ &
 $0.62_{-0.62}^{+0.97}$ & $0.68_{-0.68}^{+0.80}$ \\
& & & \\
Including \xse\ & $0.85_{-1.01}^{+0.68}$ &
 $0.24_{-0.30}^{+0.24}$ & $0.30_{-0.36}^{+0.28}$ \\
& & & \\ \hline 
\end{tabular}
\caption{Measured values of the three \tgc\ parameters using the \Qqqq\
event selection from the 183~\GeV\ data, with and without taking into 
account the systematic uncertainties. The second row gives the expected 
statistical error from Monte Carlo studies. The fourth row shows the results
after combining with the 172~\GeV\ data, and in the last row
we list the results after combining with the \xse\ information
of \Qqqq\ data from the 161, 172 and 183~\GeV\ \Com\ energies.}
\label{tab:qqqqres}
\end{center}
\end{table}
 

\section{Analysis of the \WWlnln\ angular distributions}
\label{sec:lnln}
In a \WWlnln\ event, there are at least two undetected
neutrinos. Fortunately, in the small-width, no-ISR approximation, the W
production and decay angles can still be reconstructed from the
measured charged lepton momenta, provided there are only two missing
neutrinos -- thus restricting $\ell=$e or $\mu$ for both leptons. 

\subsection{Event selection and reconstruction}
\label{subs:lnlnrec}
The selection procedure starts from the \WWlnln\ event sample used 
in the \xse\ analysis.
We then apply further cuts, mainly to suppress the contribution
of events with one or two $\tau$ leptons which cannot be reconstructed.
Only the 183~\GeV\ sample is analysed, as the statistics in the lower
energy samples are too small.

Simple cuts are applied to ensure that the momenta and
charges of the two leptons can be determined.
We require the events to have at least two reconstructed cones
containing charged particle
tracks passing the quality cuts used in the primary \Lnln\ selection.
The two highest-energy cones with charged particle 
tracks are assigned to the leptons and the charge of each cone 
is found by summing the charges of the tracks in the cone. The charges 
of the cones are required to be of different signs, or if one cone has 
a zero charge, the other must have a non-zero charge.

At this stage, 52\% of the selected sample
consists of events with at least one W\ra$\tau\nu$ decay. This fraction is 
suppressed by the following selections:
\begin{itemize}
\item The highest track momentum, $p_1$, in each lepton cone must 
exceed 23.0~\GeV, which is less by approximately twice 
the experimental resolution than the minimum momentum allowed for
a lepton from an on-shell W decay in an event with no ISR at
\roots=183~\GeV.
\item The two lepton cones must each have no more than two good
electromagnetic clusters\cite{CLUSTERS}.
\item Each of the lepton cones must be classified as an electron 
or a muon candidate, using the momentum of the most energetic
track in the cone ($p_1$), the electromagnetic calorimeter energy in the
cone ($E_{EM}$), and the hadron calorimeter energy in the cone
($E_{HC}$), as follows:
\begin{itemize}
\item electron candidate cones are required to have $E_{EM}/p_1 >
0.75$ and $E_{HC}/p_1<0.1$,
\item muon candidate cones are required to have $E_{EM}/p_1<0.1$ and
$E_{HC}/p_1<0.5$.
\end{itemize}
\end{itemize}
After applying these $\tau$ rejection cuts, the contamination from
events with  W\ra$\tau\nu$ decays falls to 10\%. 
This fraction is found to be essentially independent of the \tgc\
parameters.

In the data, 78 events are selected by the primary \Lnln\ selection.
Of these, 74 events pass the simple charge/momentum
reconstruction requirements, and 30 pass also the $\tau$
rejection selections. The purity of events with two charged leptons
(e, $\mu$ or $\tau$) and two neutrinos in this final sample is
predicted to be above 99\%, where the main background (0.13$\pm$0.07
events) comes from the reaction \ee\ra\ee\mm.
This level of background is negligible and is 
not considered further.

The lepton momenta are reconstructed using the track
momentum for muon candidate cones, and the energy seen in the
electromagnetic calorimeter for electron candidate cones.

The five characteristic angles are reconstructed event by event, in
the approximation of zero W width and no ISR.
This means that the two $\ell\nu$ systems are taken to have the W
mass
and the two W systems are 
assumed to recoil back-to-back in the laboratory frame, with a total 
energy equal to the centre-of-mass energy.
When these constraints are applied, the six observed momentum
components of the two charged leptons can be transformed into the five
angles (plus an overall azimuthal angle which is of no interest).
However, this transformation requires the solution of a quadratic
equation, and thus there are either two, or no, real
solutions for any specific event. 
The angle set \{\Cthw, $\phi^*_1$ and $\phi^*_2$\} suffers from this
ambiguity, but $\theta^*_1$ and $\theta^*_2$ can be determined from
the magnitudes of the lepton momenta alone.
In the ideal case, where the W bosons are
produced on-shell, and where there are no ISR or detector resolution
effects,
all events have two solutions: the ambiguity corresponds to a reflection
ambiguity for the two neutrinos in the plane defined by the two charged
lepton momentum vectors. 
The effects of W width, ISR and detector smearing
can move leptons to momentum configurations not allowed in on-shell W
decays yielding complex solutions for the momenta when attempting to
reconstruct the neutrinos.
These events lie preferentially in certain regions of the five-angle space,
and losing them from the analysis would introduce biases.
They can be recovered simply by taking the nearest
real solution in five-angle space, taken to be the complex neutrino
momentum solution with the imaginary part set to zero.
In these cases there is exactly one nearest real solution: the two
solutions with complex momenta are complex conjugates of
each other.

Of the 30 selected candidates, 21 have two solutions found and the other 9
have their angular information recovered using the nearest real solution
prescription. This is consistent with the Monte Carlo expectation of
8.9 events failing to have two reconstructed solutions.

\subsection{TGC analysis}
\label{subs:lnlnfit}
The fit for a TGC parameter $\alpha$ consists of minimising \LL,
defined as:
\[
\LL = -\sum_{\mathrm events} \log
\left( \frac{\frac{d\sigma}{d\Omega}(\alpha)}
{\int\frac{d\sigma}{d\Omega}(\alpha) f(\Omega)d\Omega} \right),
\]
where $\Omega$ represents the five-angle set (equivalent to phase
space for on-shell W bosons and no ISR) and $f$ represents the acceptance,
assumed to be independent of $\alpha$ (valid in the on-shell, no-ISR
case).
The differential \xse\ $d\sigma/d\Omega$ is determined using
the program of Bilenky \etal~\cite{BILENKY}.
For events with two solutions for the five angles, the average
of the differential \xses\ at the two solutions is used.

The advantage of this unbinned maximum likelihood approach over
binned techniques, as employed in the \Qqqq\ and \Qqln\ channel analyses,
is that it uses the full five-angle information without any loss
of information from binning.
The main disadvantage is that the na\"\i ve
\xse\ calculation of $d\sigma/d\Omega$ does not include many
of the effects which really occur:
specifically the effects of ISR and W width,
detector resolution, and backgrounds.
The detector and selection acceptance
is partially included through the acceptance function $f(\Omega)$
via an approximate analytical acceptance model.

It is essential to demonstrate that this simple
fit is effective in extracting couplings and that the effects not
modelled in the fit function have a relatively small influence.
The implications of unmodelled effects can be pernicious: they can
give rise to biases in the fitted couplings and in the
estimations of the fit errors.
These issues are addressed via Monte Carlo tests,
with high statistics to measure biases in the fit, and with large
numbers of simulated low-statistics experiments to calibrate the errors.

Studies are made with high-statistics Monte Carlo samples to test the
biases in the fit method.
These biases can arise from a variety of causes and are
examined step-by-step by considering
Standard Model samples with different effects included.
Biases and systematic errors are derived as described in the
following, with results shown in Table~\ref{tab:lnlnbias}.
In all cases the systematic errors are chosen so as to cover the
differences in Monte Carlo models and the full size of any bias
observed.
In most tests, events from different Monte Carlo generators gave
consistent results and the exceptions are noted.

\begin{itemize}
\item[a)] 
The effect of the primary \Lnln\ selection cuts is assessed by
fitting with the true momenta of decay leptons, both before and after the
selection cuts are applied, to fully simulated events.
This test is made with Monte Carlo W-pair events generated by 
\Excalibur, \Grace\ and \Koralw. 
The modelling of the main acceptance effects in the function $f$
reduces the bias from the acceptance.
\item[b)] 
Fits are made using the same fully simulated events, but
fitting both with the true and the reconstructed momenta. 
\item[c)]
The effect of the additional W\ra$\tau\nu$ rejection cuts on the
\Lnln\ signal ($\ell=$ e or $\mu$) is assessed by fitting to the
reconstructed simulated signal events before and after the cuts are
applied. 
\item[d)]
The \Wtnu\ background is next included, and the simulated
samples refitted. The size of the bias, averaged over the
\Excalibur, \Grace\ and \Koralw\ samples, was found to be small.
In this case, however, the different generators do give differences
which are on the edge of significance: 
\Excalibur\ and \Grace\ indicate a small positive bias to the results, 
\Koralw\ a slightly larger negative one.
The systematic error shown in Table~\ref{tab:lnlnbias} was chosen to
cover all the different predicted biases.
The modelling of $\tau$ polarisation and
decays is one area in which some of the generators are known to be
defective, so that the different biases seen are not surprising:
nonetheless we quote an error large enough to cover all
three models.
\item[e)]
Four-fermion effects were tested by fitting \Excalibur\ and
\Grace\ samples produced either including only the CC03 W-pair diagrams,
or when all four-fermion diagrams
contributing to \Lnln\ final states were employed, including the
contributions from extra final states such as 
\mm$\nu_{\rm e}\bar{\nu_{\rm e}}$. 
While the two Monte Carlo generators predict quite different
changes in accepted \xses\ when the four-fermion effects are
included, the shifts in fit results are small and compatible
with each other.
The central value of the bias was taken to be that of \Excalibur,
and is listed in Table~\ref{tab:lnlnbias}. 
The larger of the magnitude of the bias, and the difference between
the shifts seen with \Excalibur\ and \Grace, is taken as the associated 
systematic error.
\end{itemize}
\begin{table}[htb]
\begin{center}
\begin{tabular}{|ll|c|c|c|}
\hline
&Source & \dkg\   & \dgz  &  \lg\    \\ \hline 
a) &\Lnln\ selection    & --0.03$\pm$0.12 &  +0.05$\pm$0.05 &  +0.01$\pm$0.05 \\ 
b) &Detector resolution & --0.07$\pm$0.07 & --0.05$\pm$0.05 & --0.02$\pm$0.02 \\
c) &$\tau$ rejection    & --0.07$\pm$0.07 & --0.05$\pm$0.05 & --0.02$\pm$0.02\\
d) &$\tau$ background   &  +0.02$\pm$0.04 & --0.01$\pm$0.07 & --0.02$\pm$0.03 \\
e) &4-fermion effects   & --0.10$\pm$0.10 & --0.01$\pm$0.02 & --0.01$\pm$0.02\\
f) &Low statistics tests&       $\pm$0.56 &       $\pm$0.05 &       $\pm$0.03 \\
\hline
&Combined               & --0.25$\pm$0.61  & --0.07$\pm$0.14 & --0.06$\pm$0.09\\
\hline
& \LL\ correction factor&        1.17      &        2.07     &       1.39     \\
\hline
\end{tabular}
\caption{
Summary of biases, systematic errors and likelihood scale factors derived
from Monte Carlo tests for fits to the \Lnln\ angular distribution. Rows
a) to f) show biases derived from the different tests discussed in the
text, together with the associated systematic errors.
The last row shows the log-likelihood scale correction applied.
}
\label{tab:lnlnbias}
\end{center}
\end{table}

In addition to these detailed studies with Standard Model events, 
high-statistics bias tests were also made for non-standard couplings, to
ensure that the fit biases are not changing strongly with the
underlying \tgc s. These tests were done separately for W-pair and
full four-fermion samples. 
In most cases, the best fits were obtained with values consistent with
the generated couplings. Some fits, particularly for \dkg, show an 
additional minimum separated from
the correct value, and in some cases this may fit slightly better than
the generated coupling value.
In one case (\dkg=1 with the full four-fermion diagrams) the two
minima merge.
In all cases the true generated coupling was excluded by an amount
equivalent to a \LL\ interval of less than 0.25, with the present
statistics.
Incorrect second minima would in any case be removed in the full analysis
by the addition of the event rate information and that from the other
channels.

The high-statistics studies described above evaluate the biases 
intrinsic to the unbinned maximum likelihood fit method.
It is also essential for this fitting approach to calibrate the fit
errors, as they cannot be expected to be estimated free of bias.
This is done in the same way as for the other decay channels, using
many Monte Carlo subsamples with the same
statistics as the data (30 event).
A scale correction to the \LL\ function is calculated
such that in 68\% of the subsamples the correct TGC value is inside
the error interval. The scale
corrections are listed in Table~\ref{tab:lnlnbias}. 

The mean of the central values of these low-statistics fits were 
also compared, for each model, with the expected bias from the
high-statistics test. The results are consistent for \dgz\ and \lg\ but 
differ by 0.56 for \dkg. The larger of 
the differences or the statistical precision of the test is quoted
as the ``low statistics tests'' systematic error, f), in
Table~\ref{tab:lnlnbias}.

\subsection{Results for the \Lnln\ channel}
\label{subs:lnlnres}

\begin{figure}[htb]
\epsfig{file=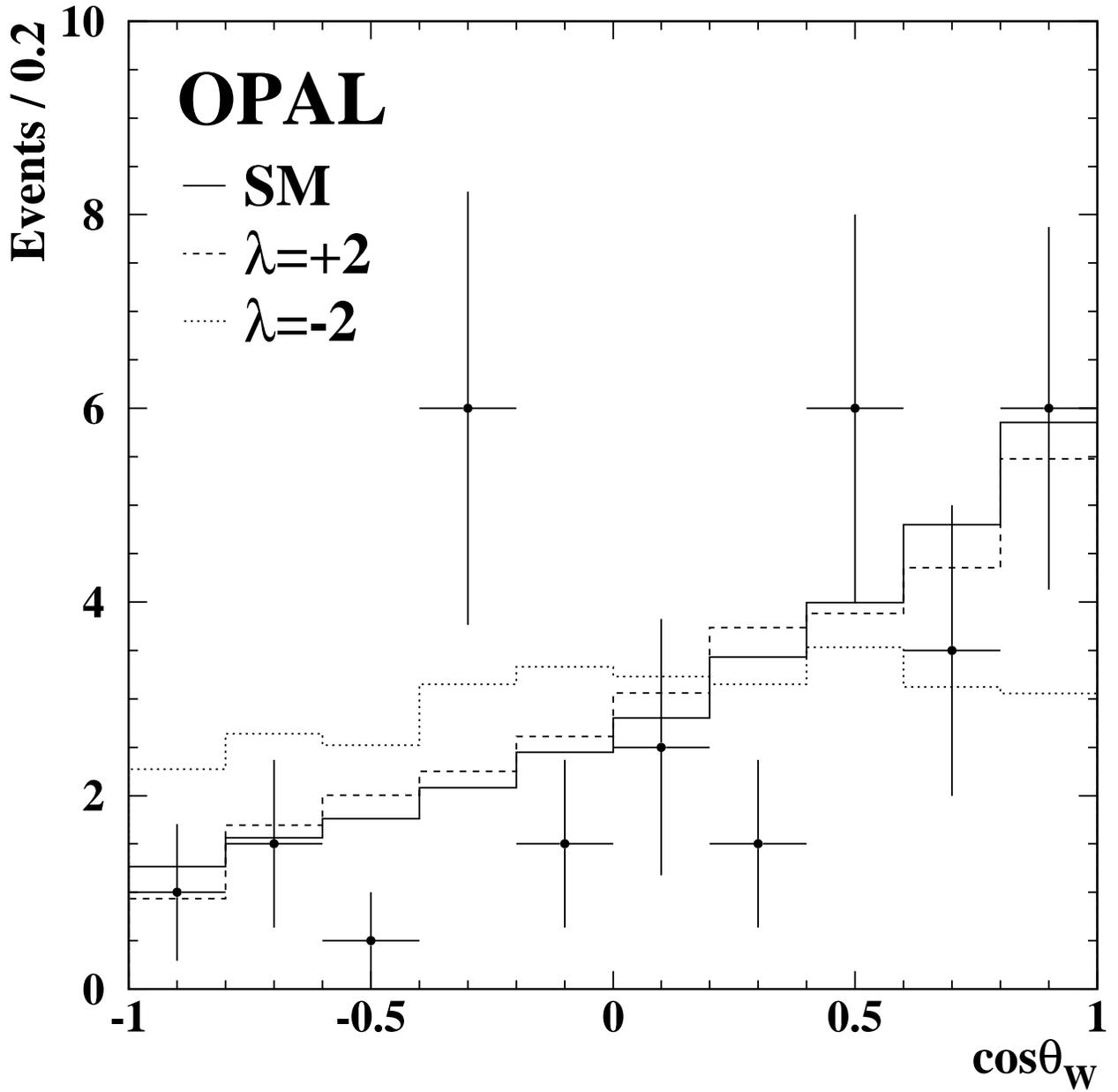,bbllx=15pt,bblly=145pt,bburx=535pt,bbury=660pt,
width=\textwidth,clip=}
\caption{\sl
Observed \Cthw\ distribution in the \Lnln\ analysis. All events enter
with a total weight of one, but in the case of events with two ambiguous
solutions for \Cthw, each solution enters with a weight of 0.5.
}
\label{fig:dacostw}
\end{figure}

\begin{figure}[htb]
\epsfig{file=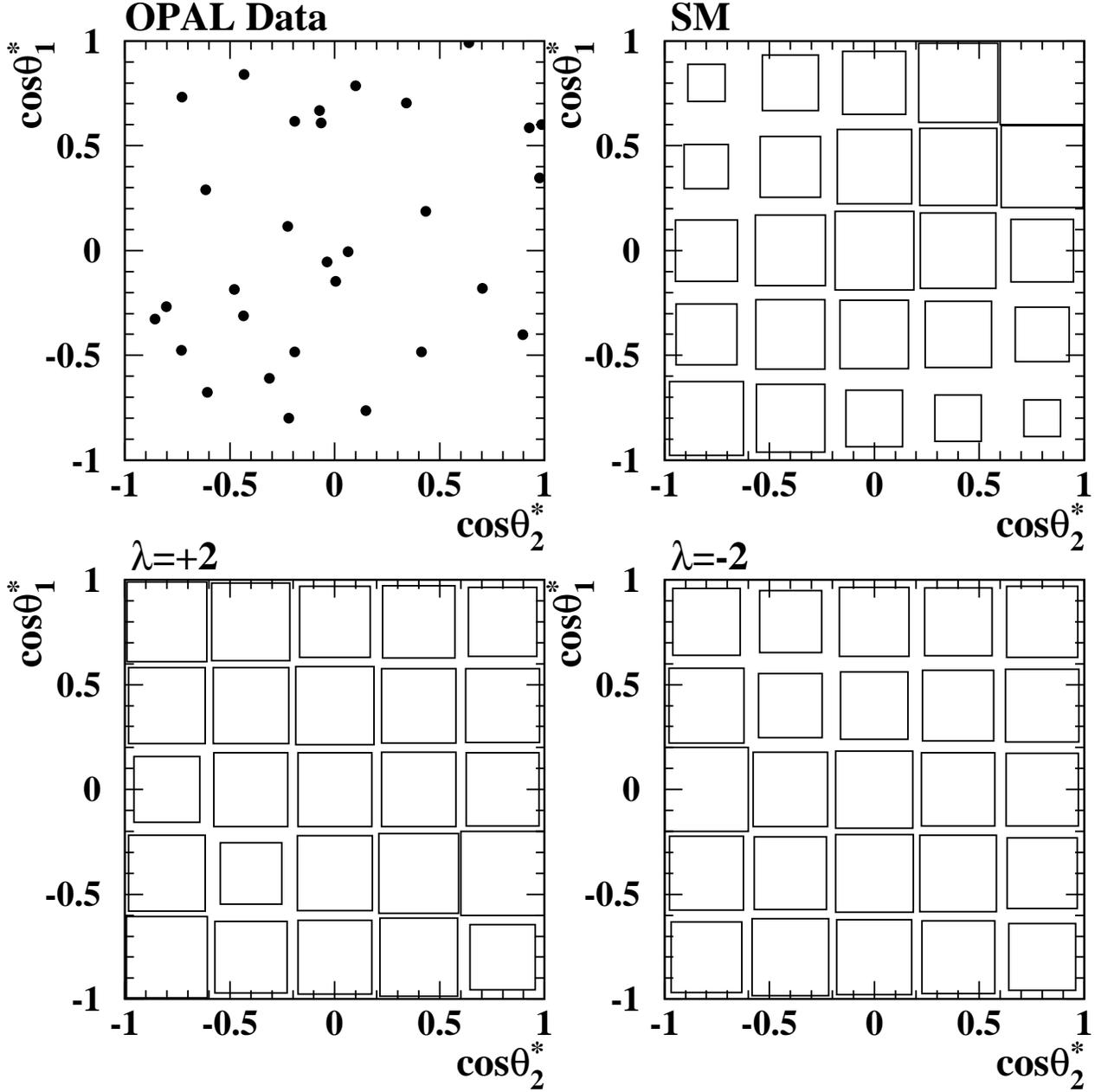,bbllx=15pt,bblly=150pt,bburx=530pt,bbury=665pt,
width=\textwidth,clip=}
\caption{\sl
Observed distribution of $\cos\theta^*_1$ vs. $\cos\theta^*_2$ in the
\Lnln\ analysis, compared to the expected distributions from samples
with standard (SM) and non-standard couplings, as indicated.
In the panels corresponding to the Monte Carlo expectations, the area
of each box is proportional to the fraction of the events falling
into that bin.
}
\label{fig:dacos12}
\end{figure}

The reconstructed \Cthw\ distribution is shown in 
Figure~\ref{fig:dacostw} and compared with the Standard Model
expectation and different \tgc\ hypotheses.
The statistics is low, but the shape is consistent with the Standard
Model expectation, as is the 
joint distribution of $\cos\theta^*_1$ and $\cos\theta^*_2$ shown
in Figure~\ref{fig:dacos12}.
In both figures there is a clear difference
visible between the different high-statistics Monte Carlo predictions.
It is interesting to note that the correlation between the two
$\theta^*_i$ angles changes with \lg. Such an effect can only be
measured well in the \Lnln\ channel because of the difficulty in
distinguishing the quark from the antiquark direction when a W boson
decays hadronically.

\begin{table}[htb]
\begin{center}
\begin{tabular}{|l|c|c|c|} \hline
Level of results      & \dkg\    & \dgz\  &  \lg\   \\ \hline 
& & & \\
Without systematics & $-1.00_{-0.92}^{+0.94}$ & $-0.79_{-0.69}^{+0.56}$
                 & $-0.29_{-0.38}^{+0.34}$ \\
& & & \\
Expected errors     & $\pm 1.08$              & $\pm 0.71$    & $\pm 0.41$ \\ 
& & & \\
Including systematics&$-0.74_{-1.16}^{+1.23}$ & $-0.76_{-0.79}^{+0.70}$
                 & $-0.23_{-0.39}^{+0.35}$ \\ 
& & & \\
Including \xse\ & $-0.30_{-0.72}^{+1.00}$ & $-0.29_{-0.31}^{+0.44}$
                 & $-0.18_{-0.30}^{+0.32}$ \\ 
& & & \\ \hline 
\end{tabular}
\caption{
Fit results at different levels, in the \Lnln\ channel.
}
\label{tab:lnlnres}
\end{center}
\end{table}

The results of the fits to the 30 selected candidates are listed
in the first row of Table~\ref{tab:lnlnres}. These results do not include
any systematic errors but they correspond to the \LL\ curves
after adjustment with the scale factor to obtain the correct 
statistical errors. The expected errors, defined to be the r.m.s. values
of the Monte Carlo subsample results, are listed in the second row of 
Table~\ref{tab:lnlnres} and they are comparable with the statistical
errors obtained from the corrected \LL\ functions.

In the next step, the \LL\ functions are 
shifted to correct for the biases and convolved with the combined
systematic uncertainties. The results are listed in the third row of
Table~\ref{tab:lnlnres}. Finally, the results are combined with those
of the \xse\ analysis corresponding to the \Lnln\ channel and
using the 161, 172 and 183~\GeV\ data samples. This combination is done
by adding the corresponding log-likelihood curves, and the results are 
listed in the last row of Table~\ref{tab:lnlnres}.   


\section{Combined TGC results}
\label{sec:combtgc}
\begin{figure}[htb]
\epsfig{file=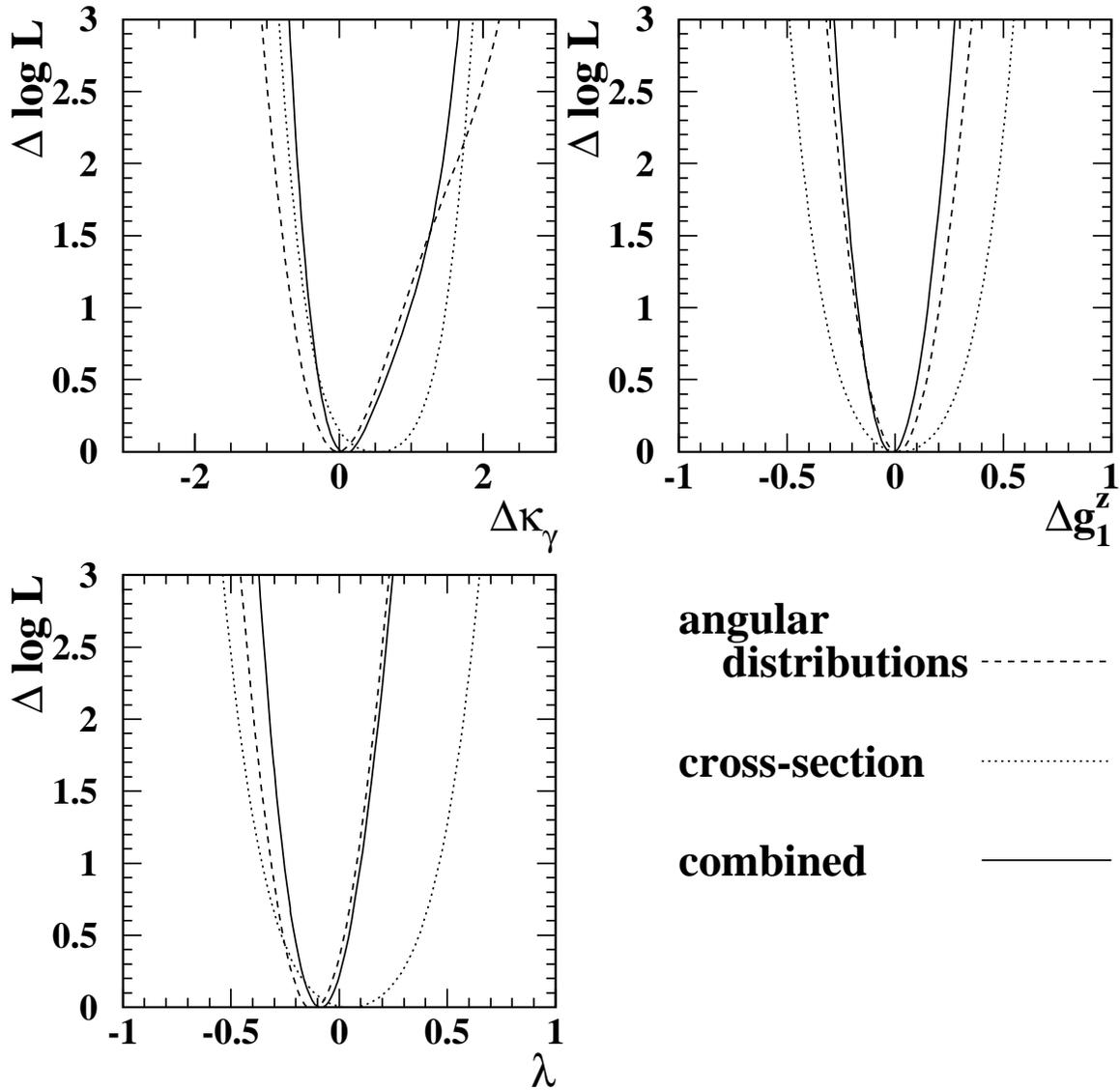,width=\textwidth}
\caption{\sl
Negative log-likelihood curves obtained using different sources
of information on the TGCs. The curves for each TGC parameter are obtained
setting the other two parameters to zero. The dashed lines are obtained
from the angular distributions, and the dotted lines from 
the total \xse. All W decay channels are used and 
systematic errors are included. The solid line is obtained by combining the
two sources of information.
}
\label{fig:rateshape}
\end{figure}

The TGC results for the three event selections using the 
angular distributions are combined by summing the corresponding 
\LL\ functions.
The correlation between the systematic errors of the three results 
is neglected, since most of the important sources of systematic errors
are relevant to a particular result and not common to all three of them.
The \LL\ functions obtained for the different couplings are shown in
Figure~\ref{fig:rateshape}. Adding these functions to those obtained
using the total \xse\ yields the combined \LL\ functions
which are plotted in Figure~\ref{fig:rateshape}. 
Table~\ref{tab:combres} lists the combined results.

\begin{table}[htb]
\begin{center}
\begin{tabular}{ |l|c|c|c| } \hline 
                      & \dkg\    & \dgz\  &  \lg\   \\ \hline
& & & \\
Combined results & \resdkg & \resdgz &  \reslam \\
& & & \\
95\% C.L. limits  & [--0.55, 1.28] & [--0.23, 0.26] & [--0.33, 0.16] \\
& & & \\ \hline 
\end{tabular}
\caption{Combined results of the three \tgc\ parameters. The result on each
parameter is obtained setting the other two parameters to zero.}
\label{tab:combres}
\end{center}
\end{table}

\begin{figure}[htb]
\epsfig{file=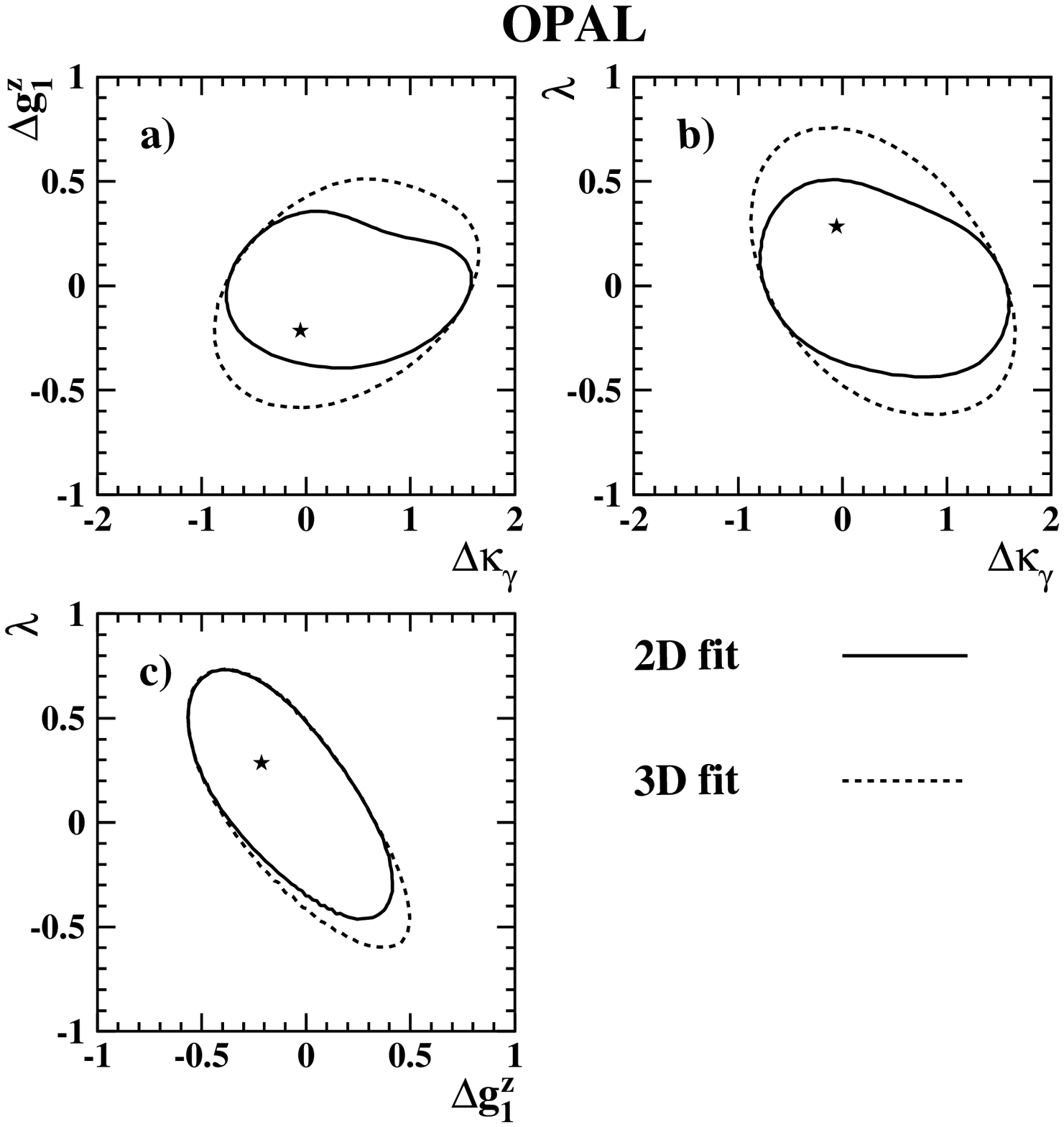,width=\textwidth}
\caption{\sl
The 95\% C.L. two-dimensional correlation contours 
for different pairs of \tgc\ parameters.
The solid lines are obtained by varying two parameters 
and fixing the third one to zero, which is  
the Standard Model value. The dashed lines show the 
projections of the three-dimensional confidence regions obtained by 
varying all three parameters. The star indicates the best 
three-parameter fit values. These results are obtained from all 
cross-section data as well as angular distributions of all \Qqln\ and
183 GeV \Qqqq\ data.
}
\label{fig:2d3d}
\end{figure}

To study correlations between the three TGC parameters we also extract the
\LL\ as a function of all three variables, \dkg, \dgz\ and \lg.
Figure~\ref{fig:2d3d} shows the 95\% C.L. contour plots obtained from 
two-dimensional fits, where the third parameter is fixed at its
\SM\ value of zero. We also perform a three-dimensional fit, where all three 
couplings are allowed to vary simultaneously. The corresponding projections 
are plotted as dashed contour lines in Figure~\ref{fig:2d3d}. As can be 
seen, the allowed range for each parameter is extended when the constraints 
on the other two parameters are removed.

These three-dimensional fits can be used to obtain results for any other set
of TGC parameters, as long as relations~(\ref{su2u1}) between the five
couplings are satisfied. For example, we determine the $\alpha$ parameters
used in our previous publications~\cite{tgc161-analysis,tgc172-analysis}
to be \abf=\resdkg, \awf=\resawf\ and \aw=\reslam. The result on each 
$\alpha$ parameter is obtained assuming that the other two parameters 
vanish.


\section{Summary and conclusions}
\label{sec:summary}  
Using a sample of 877 \WW\ candidates collected at \LepII\ at a \Com\ energy
of 183~\GeV, we measure the total \xse\ of,
$$ \sigma(\eeWW) = \measxs \pm \sigst \pm \sigsys \; {\mathrm{pb}}$$
under the assumption that the W boson decay
branching fractions and distributions of production and decay angles are
all according to the \SM\ expectations.

A measurement of the W branching fractions is made using the combined
161~\GeV, 172~\GeV\ and 183~\GeV\ data samples. The results for the
different leptonic decay channels are consistent with each other, as 
expected from lepton universality. Assuming \SM\ total \xse\ and lepton 
universality, we obtain the hadronic decay fraction to be 
($\measbr\pm\brst\pm\brsys$)\%. From this result, a value for the CKM 
mixing matrix elements $\Vcs=0.99 \pm 0.06 \pm 0.02$ is extracted, 
using also the measurements of the other matrix elements not involving the 
top quark.

The total \xse\ measurement, being consistent with the \SM\ 
prediction of \GENTxs~\Ipb, can be used to place limits on anomalous
triple gauge boson couplings. Those couplings are also investigated in 
an independent way, using the W-pair production and decay angular 
distributions
for \Qqln, \Qqqq\ and \Lnln\ final states. The \WWqqln, being the most
sensitive channel for this study, is analysed by three different methods,
one of them utilising for the first time the spin density matrix of the 
W decay. All three methods lead to consistent results. The spin density
matrix is used to measure the \xses\ to produce transversely
and longitudinally polarised W bosons. Integrating over all angles, the
fraction of longitudinally polarised W bosons is determined to be
0.242$\pm$0.091$\pm$0.023.   

The TGC measurements for all decay channels are combined, and 
the results obtained are,
\begin{eqnarray*}
\dkg & = & +\resdkg, \\     
\dgz & = & +\resdgz, \\    
\lg  & = & \reslam,    
\end{eqnarray*}
where each parameter is determined setting the other two parameters to
zero.
The precision of these results is comparable to the latest D0 results 
obtained from boson pair production at the Tevatron \pbp\ collider~\cite{D0}.
These results supersede those from our previous 
publications~\cite{tgc161-analysis,tgc172-analysis}.
They are all consistent with the \SM\ value of zero.


\section*{Acknowledgements}
We thank T. Wiseman for his contribution to this analysis. \\
We particularly wish to thank the SL Division for the efficient operation
of the LEP accelerator at all energies
 and for their continuing close cooperation with
our experimental group.  We thank our colleagues from CEA, DAPNIA/SPP,
CE-Saclay for their efforts over the years on the time-of-flight and trigger
systems which we continue to use.  In addition to the support staff at our own
institutions we are pleased to acknowledge the  \\
Department of Energy, USA, \\
National Science Foundation, USA, \\
Particle Physics and Astronomy Research Council, UK, \\
Natural Sciences and Engineering Research Council, Canada, \\
Israel Science Foundation, administered by the Israel
Academy of Science and Humanities, \\
Minerva Gesellschaft, \\
Benoziyo Center for High Energy Physics,\\
Japanese Ministry of Education, Science and Culture (the
Monbusho) and a grant under the Monbusho International
Science Research Program,\\
Japanese Society for the Promotion of Science (JSPS),\\
German Israeli Bi-national Science Foundation (GIF), \\
Bundesministerium f\"ur Bildung, Wissenschaft,
Forschung und Technologie, Germany, \\
National Research Council of Canada, \\
Research Corporation, USA,\\
Hungarian Foundation for Scientific Research, OTKA T-016660, 
T023793 and OTKA F-023259.
 
 
\bibliography{pr260}

\begin{thebibliography}{10}

\bibitem{LEP2YR}
Physics at LEP2, edited by G.\ Altarelli, T.\ Sj{\"{o}}strand and F.\ Zwirner,
  CERN 96-01 Vol. 1, 525.

\bibitem{HAGIWARA}
K.\ Hagiwara, R.D.\ Peccei, D.\ Zeppenfeld and K.\ Hikasa,
  \NPB{282}{1987}{253}.

\bibitem{BILENKY}
M.\ Bilenky, J.L.\ Kneur, F.M.\ Renard and D.\ Schildknecht,
  \NPB{409}{1993}{22}; \NPB{419}{1994}{240}.

\bibitem{GAEMERS}
K.\ Gaemers and G.\ Gounaris, \ZPC{1}{1979}{259}.

\bibitem{DERUJULA}
A.\ De Rujula, M.B.\ Gavela, P.\ Hernandez and E.\ Masso, \NPB{384}{1992}{3}.

\bibitem{HISZ}
K.\ Hagiwara, S.\ Ishihara, R.\ Szalapski and D.\ Zeppenfeld,
  \PLB{283}{1992}{353}; \PRD{48}{1993}{2182}.

\bibitem{CDF}
CDF Collaboration, F.\ Abe \etal, \PRL{78}{1997}{4536}.

\bibitem{D0}
D0 Collaboration, B.\ Abbott \etal, \PRD{58}{1998}{31102}.

\bibitem{tgc161-analysis}
\opalacker, \PLB{397}{1997}{147}.

\bibitem{OTHERLEP161-tgc}
DELPHI Collaboration, P.\ Abreu \etal, \PLB{397}{1997}{158}; \\ L3
  Collaboration, M.\ Acciarri \etal, \PLB{398}{1997}{223}.

\bibitem{tgc172-analysis}
\opalacker, \EPC{2}{1998}{597}.

\bibitem{OTHERLEP172-tgc}
L3 Collaboration, M.\ Acciarri \etal, \PLB{413}{1997}{176}; \\ DELPHI
  Collaboration, P.\ Abreu \etal, \PLB{423}{1998}{194}; \\ ALEPH Collaboration,
  R.\ Barate \etal, \PLB{422}{1998}{369}.

\bibitem{SEKULIN}
R.L.\ Sekulin, \PLB{338}{1994}{369}.

\bibitem{OPAL}
OPAL Collaboration, K.\ Ahmet \etal, \NIMA{305}{1991}{275}; \\ S.\ Anderson
  \etal, \NIMA{403}{1998}{326}.

\bibitem{SW}
B.E.\ Anderson \etal, IEEE Transactions on Nuclear Science, \textbf{41} (1994)
  845.

\bibitem{GENTLE}
D.\ Bardin \etal, Nucl. Phys. B, Proc. Suppl. {\bf 37B} (1994) 148; \\ D.\
  Bardin \etal, ``GENTLE/4fan v.\ 2.0: A Program for the Semi-Analytic
  Calculation of Predictions for the Process \epem\ra 4f'', DESY 96-233,
  hep-ph/9612409.

\bibitem{MWPDG}
D0 Collaboration, B.\ Abbott \etal, \PRL{80}{1998}{3008}; \\ CDF Collaboration,
  F.\ Abe \etal, \PRL{75}{1995}{11}.

\bibitem{GOPAL}
J.\ Allison \etal, \NIMA{317}{1992}{47}.

\bibitem{KORALW}
M.\ Skrzypek \etal, \CPC{94}{1996}{216}; \\ M.\ Skrzypek \etal,
  \PLB{372}{1996}{289}.

\bibitem{EXCALIBUR}
F.A.\ Berends, R.Pittau and R.\ Kleiss, \CPC{85}{1995}{437};\\ F.A.\ Berends
  and A.I.\ van Sighem, \NPB{454}{1995}{467}.

\bibitem{GRC4F}
J.\ Fujimoto \etal, \CPC{100}{1997}{128}.

\bibitem{PYTHIA}
T.\ Sj{\"{o}}strand, \CPC{82}{1994}{74}.

\bibitem{HERWIG}
G.\ Marchesini \etal, \CPC{67}{1992}{465}.

\bibitem{KORALZ}
S.\ Jadach \etal, \CPC{79}{1994}{503}.

\bibitem{BHWIDE}
S.\ Jadach, W.\ Placzek and B.F.L.\ Ward, \PLB{390}{1997}{298}.

\bibitem{PHOJET}
R.\ Engel, \ZPC{66}{1995}{203}; \\ R.\ Engel and J.\ Ranft,
  \PRD{54}{1996}{4244}.

\bibitem{VERMASEREN}
J.A.M.\ Vermaseren, \NPB{229}{1983}{347}.

\bibitem{mass172-analysis}
\opalacker, \EPC{1}{1998}{395}.

\bibitem{acopll-analysis}
\opalacker, \EPC{4}{1998}{47}.

\bibitem{fpair183-analysis}
\opalacker, CERN-EP/98-108, to be published in Eur. Phys. J. C.

\bibitem{TKMH}
\opalalexander, \ZPC{52}{1991}{175}.

\bibitem{DURHAM}
N.\ Brown and W.J.\ Stirling, \PLB{252}{1990}{657}; \\ S.\ Catani \etal,
  \PLB{269}{1991}{432}; \\ S.\ Bethke, Z.\ Kunszt, D.\ Soper and W.J.\
  Stirling, \NPB{370}{1992}{310}; \\ N.\ Brown and W.J.\ Stirling,
  \ZPC{53}{1992}{629}.

\bibitem{GCE}
\opalakrawy, \PLB{253}{1990}{511}.

\bibitem{QCDMATRIX}
S.\ Catani and M.H.\ Seymour, \PLB{378}{1996}{287}.

\bibitem{ERT}
R.K.\ Ellis, D.A.\ Ross and A.E.\ Terrano, \NPB{178}{1981}{421}.

\bibitem{COSNR}
O.\ Nachtmann and A.\ Reiter, \ZPC{16}{1982}{45}; \\ M.\ Bengtsson,
  \ZPC{42}{1989}{75}.

\bibitem{Karlen}
D.\ Karlen, physics/9805018, Computers in Physics {\bf 12} (1998) 380.

\bibitem{qqqqQCD}
\opalalexander, \ZPC{69}{1996}{543}.

\bibitem{pdg}
C.\ Caso \etal, \EPC{3}{1998}{1}.

\bibitem{opalmw161}
\opalacker, \PLB{389}{1996}{416}.

\bibitem{OO}
M.\ Davier, L.\ Duflot, F.\ LeDiberder and A.\ Rouge, \PLB{306}{1993}{411}; \\
  M.\ Diehl and O.\ Nachtmann, \ZPC{62}{1994}{397}.

\bibitem{BARLOW}
R.\ Barlow and C.\ Beeston, \CPC{77}{1993}{219}.

\bibitem{GOUNARIS}
G.\ Gounaris, J.\ Layssac, G.\ Moultaka and F.M.\ Renard,
  \IJMP{19}{1993}{3285}.

\bibitem{BEC}
L.\ L{\"{o}}nnblad and T.\ Sj{\"{o}}strand, \EPC{2}{1998}{165}.

\bibitem{RECONN}
T.\ Sj{\"{o}}strand and V.A.\ Khoze, \ZPC{62}{1993}{281}.

\bibitem{ARIADNE}
L.\ L{\"{o}}nnblad, \CPC{71}{1992}{15}.

\bibitem{CLUSTERS}
\opalalexander, \ZPC{72}{1996}{191}.

\end{thebibliography}


\end{document}